Chapter 14

# Moist Convective Storms on Saturn

Agustín Sánchez-Lavega[1], Georg Fischer[2], Cheng Li[3],
Enrique García-Melendo[4], Teresa del Rio-Gaztelurrutia[1]

[1]Universidad del País Vasco UPV/EHU, Bilbao, Spain
[2]Austrian Academy of Sciences, Graz, Austria
[3]University of California, Berkeley, USA
[4] Universitat Politècnica de Catalunya UPC, Terrassa, Spain

**Abstract**

Convective storms (CS) on Saturn appear as bright clouds, irregular in shape and rapidly changing over time, accompanied by lightning phenomena. They cover a wide range of size scales, from 500 km or less and up to sizes that encircle the planet. In this chapter we focus on medium or synoptic-scale events (typical sizes ~ 2,000 - 7,000 km) and planetary-scale major storms known as Great White Spots (GWS). In general, the storms emerge in westward jets and in the peak of the strong eastward equatorial jet. They have been observed preferentially in the northern hemisphere and seem to occur in the same latitude in intervals of 30 and 60 years coinciding with the times of maximum insolation within the seasonal cycle. Moist convection models locate the origin of the storms in the water clouds (at about 10 bars of pressure) and to some extent in ammonia clouds, and their thermodynamics explain the possible periodicity of the phenomena. Once triggered, the storms grow and expand, and their interaction with the zonal wind system gives rise to anticyclonic ovals and to disturbances that propagate zonally covering an entire latitude band in the case of the GWS.

## 14.1 Introduction

According to thermochemical models, there are at least three cloud layers (ammonia, ammonium hydrosulfide and water) extending vertically in the upper troposphere of Saturn, between approximately the pressure levels of 1 bar and 10 bar (Weidenshilling and Lewis, 1973; Atreya et al., 1999). The clouds form from the condensation of these compounds with the upper cloud decks expected to be composed of ice particles and liquid droplets expected deep within the water clouds, and the release of latent heat warms moist air parcels and can lead to ascending motions in the atmosphere. If these updrafts are massive and the ascent vigorous, they form convective storms, with cloud-tops that can reach the upper troposphere, becoming visible (Sánchez-Lavega and Battaner, 1986). Thus, convective storms manifest themselves at visual and near infrared wavelengths as bright spots rapidly changing over time. However, the amount of ground-based observations of features in Saturn's atmosphere is not large (as compared to Jupiter), due both to the small size of the planet as seen in the sky (about 18 arcsec) and to the low contrast that these features present over the bland aspect of the planet, caused by the presence of hazes (West et al., 2009). Historically, there were few reports of spots on Saturn (Sánchez-Lavega, 1982) and only Great White Spots (GWS), unusual phenomena of massive convection, were studied with some detail with telescopes (Sánchez-Lavega, 1994; Sánchez-Lavega et al., 2018). The entry of digital detectors into operation in the 1980s, and the recent development in last few years of "lucky imaging" (see e. g. Mendikoa et al., 2016), changed the situation, and nowadays small telescopes are able to detect different types of spots and features, including convective storms, on the planet. However, our detailed knowledge of the atmospheric dynamics and in particular of storms in Saturn's atmosphere comes essentially from observations during the Voyager 1 and 2 flybys in 1980-81 (Smith et al., 1981, 1982; Ingersoll et al., 1984) and from those of the Cassini mission between 2004 and 2017 (Ingersoll, 2020). Added to these are the observations made occasionally with the Hubble Space Telescope (HST), which has provided high-quality images of some of these phenomena since its entry into operation in 1990. Unless otherwise specified, throughout this chapter we use planetocentric latitudes and System III longitudes as the frame of reference for the planet's rotation.

## 14.2 Observations of Convective Storms

We refer to a spot or feature in Saturn's atmosphere as a convective storm (CS) when we observe the sudden outbreak of a bright cloud (at optical and near infrared wavelengths) that shows an irregular morphology, and grows in size and changes or expands in a brief period of time. Depending on its size, a storm can last a few days to months until the activity ceases and the clouds that make up the convective complex finally dissipate. Convective storms can modify temperatures and chemical compositions at different altitudes, as was observed in detail in the case of 2010 GWS (Sánchez-Lavega et al., 2018). It developed electrical phenomena that give rise to lightning, detected at optical wavelengths and in radio emissions, as described in Section 14.3.

We classify Saturn's convective storms by their size, morphology, and evolution as observed in visual images of the cloud field. Accordingly, storms can be broadly grouped in three categories following their spatial extent and temporal duration (Sánchez-Lavega et al., 2019). First, we have mesoscale phenomena with a cloud size of ~ 500-1000 km or smaller. These storms are beyond the detection of telescopes on Earth, and only the largest are within the limit of visibility of the HST, and thus will not be discussed in detail. We know of their existence thanks to the high-resolution images from Voyager 1 and 2 and the Cassini spacecraft, which revealed their presence in many regions of the planet. A good example is the North Polar Region, where clusters of small storms have been observed in a large area (Antuñano et al., 2018). A second type of storms develops at intermediate or synoptic-scale, where the active region of bright clouds reaches sizes of ~ 2,000 – 8,000 km. These storms undergo rapid morphology changes, and associated clouds are sometimes expanded zonally by the prevailing ambient winds, but not fully encircling the planet. These storms can be observed with ground-based telescopes and with HST, and the Voyagers and Cassini have allowed their study in great detail. Sections 14.2.2 and 14.2.3 are dedicated to them. Finally, we have the infrequent major planetary-scale events known as Great White Spots (GWS) (Sánchez-Lavega et al., 1994; Sánchez-Lavega et al., 2018). In these mighty storms, the active region (known as the head) can reach ~ 10,000 km or more after its growth and expansion, and the vigorous dynamics of the storm propagates from the head in the zonal direction, forming a disturbed band that fully encircles the planet. A rich set of dynamics accompanies these events, which will be described in section 14.2.3.

### 14.2.1 Synoptic-scale storms

In the following sections we use the letters $u$, $v$, and $w$ to denote storm velocities. The letter $u$ denotes longitudinal speeds, with $u>0$ for eastward and $u<0$ for westward speeds. The letter $v$ denotes latitudinal speeds, which are positive when the movement is poleward. Finally, $w$ stands for vertical speeds being positive for upward movements.

Convective storms of synoptic scale have been observed at different latitudes and epochs. During the Voyager 2 flyby a disperse system of plume-like bright features was observed from latitudes $\varphi_c \sim 2.5°N$ to $5°N$ ($\varphi_c$ is the planetocentric latitude), moving with velocities of the peak of the equatorial jet $u \sim 455 – 465$ ms$^{-1}$ (Sánchez-Lavega et al., 2000). The shape of these features was reminiscent of convective phenomena. Nevertheless, at the Voyager's epoch, most of the storm activity was located in a narrow band in the anticyclonic side of a westward jet (poleward of the jet in the Northern Hemisphere) with zonal velocity $u = -20$ ms$^{-1}$, between latitudes $\varphi_c \sim 33.5°N$ - $38°N$ (Smith et al., 1981, 1982; Ingersoll et al., 1984), the same latitude where the GWS 1903 and 2010 events occurred. During the Voyagers flybys, there were intermittent eruptions of short-lived white bright storms, irregular in shape, typically of size 2,000-3,000 km (Figure 1a-b) (Sromovsky et al., 1983). Hunt et al. (1982) studied these in Voyager 1 images and estimated from the growing size of the clouds a flow divergence of $5x10^{-5}$ s$^{-1}$, which, using mass conservation, implied vertical velocities at cloud tops of $w \sim 1$ ms$^{-1}$. During the Voyager 2 flyby, some of these storms formed in the wake of anticyclonic vortices known as Brown Oval spots (Smith et al. 1981, 1982; García-Melendo et al., 2007) generating turbulent regions denoted as "v-shaped features" (Sromovsky et al., 1983). The activity probably lasted throughout 1983 according to ground-based imaging (Suggs and Beebe, 1983). From 2002 until 2010, similar storm activity was observed to be concentrated in a symmetrical band in the southern hemisphere of the planet, informally known as the "storm alley", which is described in detail in section 14.2.2.

Convective events of synoptic-scale have also been observed at subpolar and polar latitudes. In 1994 a bright spot with a size of $\sim$ 5,000 - 7,000 km was observed in the southern hemisphere at $\varphi \sim 50°$S in a westward jet with a velocity of 15 ms$^{-1}$ (Sánchez-Lavega et al., 1994). The clouds from the storm expanded eastward with a velocity $u = 30$ ms$^{-1}$ in System III reaching a distance of $\sim$ 20,000 km from the head. The measured velocities were consistent with the zonal wind velocity profile at these latitudes (Sánchez-Lavega et al., 2000; García-Melendo et al., 2011).

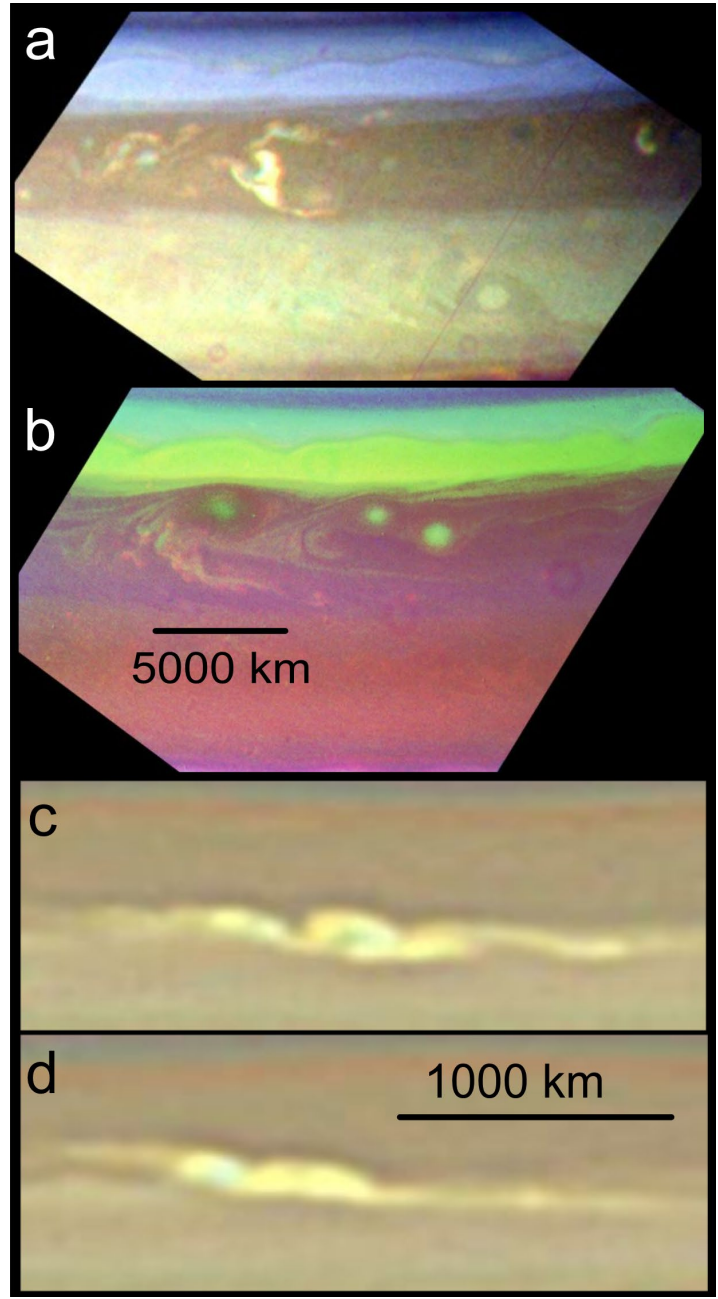

**Figure 14.1.** *Synoptic-scale convective storms on Saturn: (a-b) Mid-latitude northern hemisphere features at 35°N observed by Voyager 1 (a) and Voyager 2 (b) (from Smith et al., 1982). In (a) the convective storm is the bright and irregular morphology structure, and in (b) the anticyclonic vortex BS1 (dark oval with bright spot at center) and other structures in the active convective area are shown. (c-d) Polar storms WS1 and WS2 observed with the Hubble Space Telescope in 2018 (from Sánchez-Lavega et al., 2020). The horizontal bar shows the scale of the image.*

In 2018, four synoptic storms (called WS1, WS2, WS3, WS4) with east–west lengths of ~ 6,000 km formed sequentially in the North Polar Region at close latitudes ($\varphi_c$~ 62.1°N, 65.3°N, 68.2°N, 70.9°N respectively). They experienced mutual encounters, and the combined effect of the storms led to zonal disturbances that affected a full latitude band from 62.0°N to 71°N (north-south width of ~ 9,000 km) (Figure 14.1 c-d, Sánchez-Lavega et al., 2020). The first storm, which erupted in a pre-existing cyclonic vortex, was the most enduring (214 days), moved with the winds with $u = 60$ ms$^{-1}$ and migrated poleward with a velocity $v$ ~ 0.5 ms$^{-1}$. The second outbreak occurred northwards, moving at $u$ ~ 8 ms$^{-1}$. The third and fourth storm erupted on the cyclonic and anticyclonic side of the westward jet centered at $\varphi c \sim 69.4°N$, respectively, and both moved according to the zonal wind profile at $u \sim -5$ ms⁻¹. The time intervals between different outbreaks were 56 days (WS1-WS2), 25 days (WS2-WS3) and 56 days (WS3-WS4). The initial separation between storms ranged from $\sim 6,700 – 14,000$ km, and all of them drifted poleward with a meridional velocity $v \sim 0.5 – 1.1$ ms⁻¹.

Following a year of calm, a new episode took place in March-May 2020, further north than the 2018 storms. The first main storm erupted at 72.5°N and spanned 4,500 km. It developed a tail that extended zonally $\sim 40,000$ km (Sánchez-Lavega et al., 2021). During its lifetime of $\sim$ 40 days, the storm migrated poleward with a velocity $v \sim 1.2$ ms⁻¹ reaching latitude 73.6°N. A new outbreak of two distinct storms occurred 37 days later, at latitude 73.3°N in the hexagon's south edge, separated by 2,250 km and 9,250 km from the first storm. All three storms emerged on the anticyclonic side of the hexagon jet, and their clouds penetrated into the southern side of the hexagonal wave and its intense eastward jet.

The 2018 and 2020 polar storms took place in locations with different zonal wind maxima and different ambient vorticities, and they did not produce a noticeable change in the zonal wind profile. Although the hexagon shape was perturbed after the storm, the wave did not undergo any change either in its rotation period, that matches System III, or in its embedded eastward jet. This result reinforces the idea that Saturn's hexagon is a well rooted structure with a possible direct relationship with the bulk rotation of the planet (Sánchez-Lavega et al., 2021). The temporal cadence and close separation in longitude of the outbreaks suggest that a triggering instability existed, probably based within the water clouds (P $\sim$ 10 bar).

### 14.2.2 Southern mid-latitude "storm alley"

From 2002 until the eruption of the 2010 GWS, all storm-activity in Saturn's atmosphere was concentrated in the "storm alley", a $\sim 1.5°$-wide latitude band centered a at 36.2°S, in the anticyclonic side of a weak westward jet (Porco e a., 2005; Vasavada et al., 2006; Dyudina et al., 2007) (Figure 14.2). The presence of convective activity was captured initially in ground-based and HST images from 2002 to 2004 (Sánchez-Lavega et al., 2003, 2004a). The arrival of Cassini in 2004 showed the features to be similar to those observed by the Voyagers, and they were soon characterized as storms due to their brightness, rapid cloud evolution and correlation with radio emissions from lightning, the so-called Saturn Electrostatic Discharges (SEDs) to a collocated origin with these cloud features (see section 14.3.2). In the Cassini era, temporal sampling was completed with low resolution images of amateur astronomers, whose joint effort allowed a continuous monitoring of the planet.

The clouds associated with these synoptic storms can extend laterally up to $\sim 4000$ km, and are quite distinctive in Cassini Imaging Science Subsystem (ISS) high-resolution images (Figure 14.2). In broadband filters, they appear as bright compact areas with irregular shapes that change significantly in a time scale of hours. Storm clouds are less conspicuous in ISS MT2 methane filters, and undetectable in MT3 and UV filters, which sound the upper regions of the troposphere (Dyudina et al., 2007).

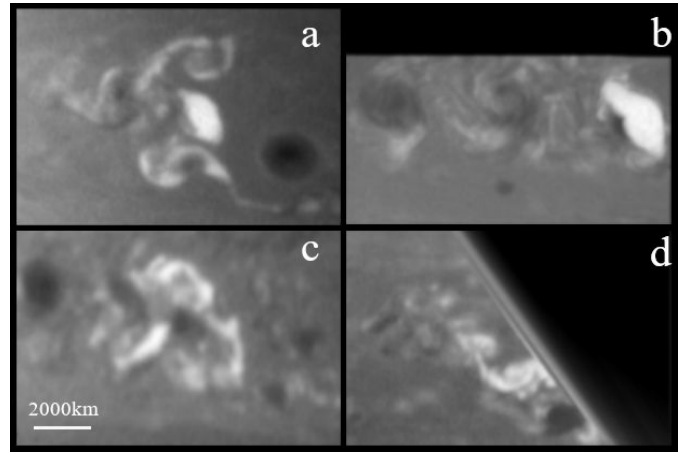

**Figure 14.2.** *Several examples of synoptic-scale storms along the period of activity in the storm alley. Projections cover 15° longitude and 8° latitude, centered at 36° planetocentric latitude. Scale is indicated with a white line. Panel a: 13 Sep 2004; Panel b: 15 Feb 2008: Panel c: 4 Mar 2008; Panel d: 27 Jan 2009. Original images for panels a and b were taken with the ISS Narrow Angle Camera and CB2 CL1 filters, and for panels c and d, with the ISS Wide Angle Camera and CB3 CL2 filters.*

The activity in the storm alley was episodic, with periods with frequent outbursts and long periods where both Earth and Cassini observers did not detect any storm. A summary of all events, together with their longitudinal drifts, is presented in Figure 14.3. There were three relatively brief episodes, two in 2004 and one 2006, each lasting a couple of months, a long episode in the first semester of 2008 (which started in late 2007) and quasi-continuous activity after 2009 (with a precursor in late 2008), with a brief gap at the beginning of 2010. The same pattern was in principal observed for SEDs (Fig14.8), therefore excluding the potential bias that could have been caused by a lack of Cassini ISS observations during Saturn opposition. Only the brief episode in early 2004 could not be detected by the radio wave instrument, as Cassini was still too far from Saturn for the detection of SEDs. Except for the events in 2004 and 2006, ISS temporal sampling was sparse, and the contribution of images from Earth was essential to identify different ISS images as belonging to the same episode. An analysis of the average drift velocities of each episode indicates that they all move slightly slower than the ambient winds, as shown in Figure 14.3.

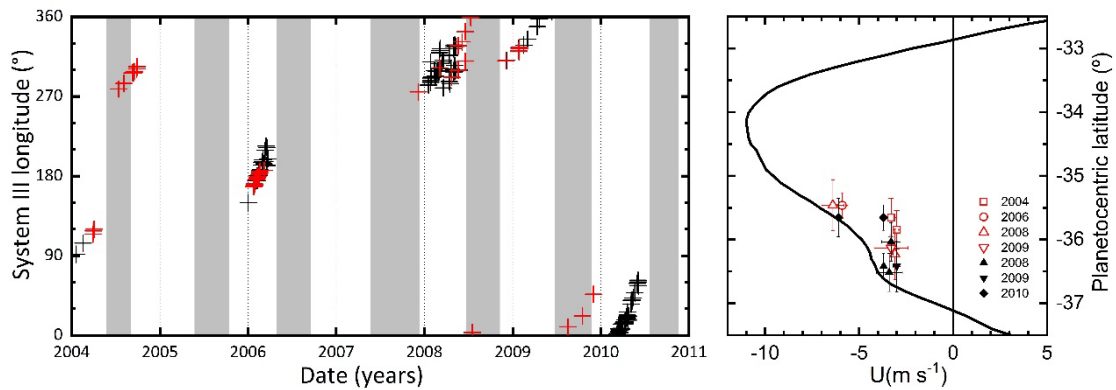

**Figure 14.3.** *Left panel: System III sources of storm activity in the storm alley. Red crosses indicate Cassini ISS observations, black crosses are observations from Earth. Grey shaded areas represent Saturn opposition. Right panel: Drift velocity of different episodes, compared with the zonal wind profile of Garcia-Melendo et al. (2011). Red and black symbols indicate velocities derived from linear fits of Cassini ISS data and observations from Earth, respectively.*

The episode of autumn 2004 (known as the "dragon storm") was analyzed in great detail using ISS images (Dyudina et al., 2007). It revealed the presence of a source of convective activity that triggered intermittently, and gave rise to a number of anticyclones (Figure 14.4). The source, located at planetographic 35.8°S latitude moved with -2.8 ms$^{-1}$, slower than the ambient zonal winds. Each outburst was short, lasting less than 24 hours with area growth rates of ~ 23 km$^2$s$^{-1}$. After the eruptions, the storm's bright clouds expanded following irregular patterns, and finally generated anticyclonic vortices of very similar shape and size. These vortices migrated to lower latitudes and moved faster at -8.8 ms$^{-1}$, closer to ambient velocity. Anticyclones appear dark in CB filters and bright at short wavelengths (Figure 14.2), similar to those observed in the other hemisphere during the Voyager epoch. Although the other episodes have sparser temporal sampling, they confirm this kind of evolution. For example, the poorer coverage of the storm of 2006 reveals the same pattern of several relatively brief outbursts followed by evolution into vortices merges (Dyudina et al., 2007).

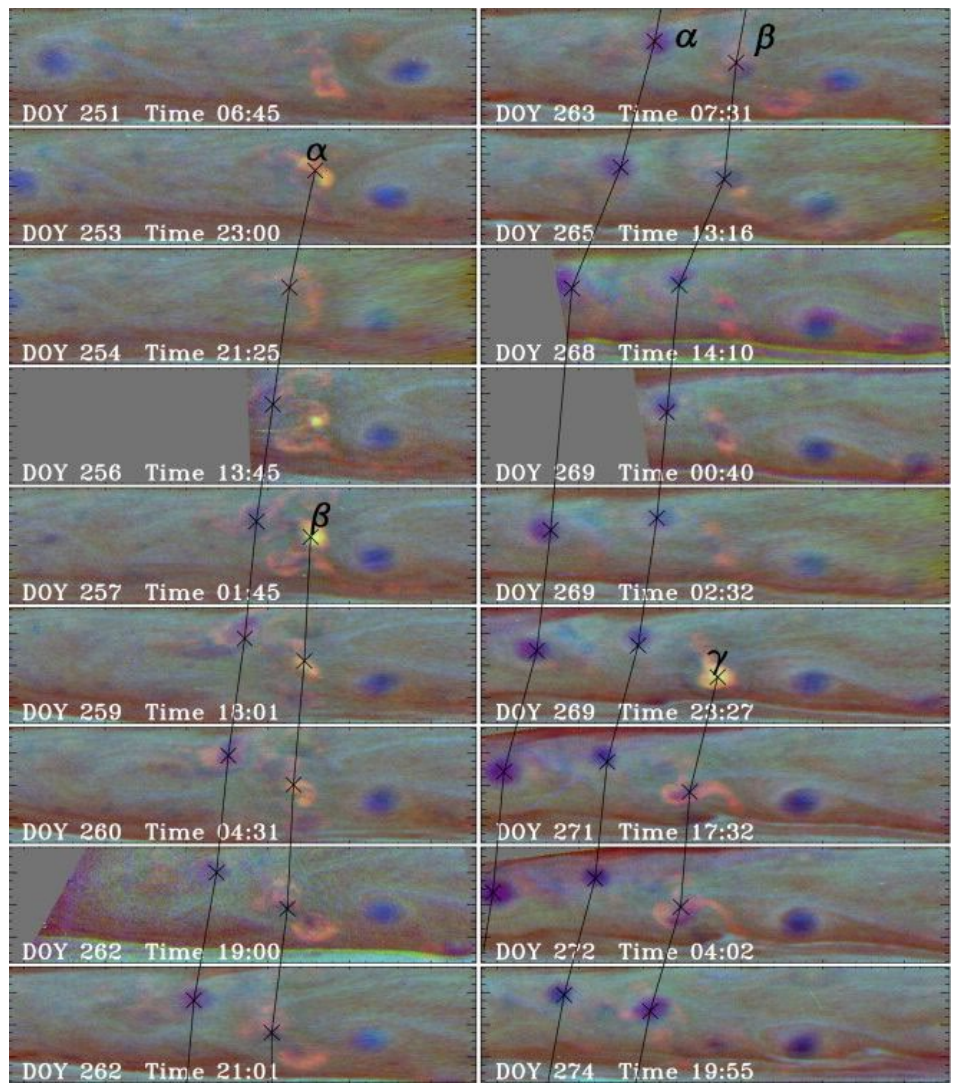

**Figure 14.4.** *Activity in the storm alley in September 2004. Red-Green-Blue channels correspond to 750 nm, 727 nm and 889 nm, respectively. The lines marked α, β and γ represent the evolution of individual eruptions into vortices. Maps are centered at planetocentric 36°S latitude and extend 7° in latitude and 23° in longitude. From Dyudina et al. (2007).*

While the Cassini temporal sampling of the 2008 "two-cell storm" event (Fischer et al., 2019) is scarcer, SED detections and frequent observations from Earth indicate almost continuous activity. Again, the morphology and size of the detected events are similar to those of the 2004 episode. The storm-to-anticyclone evolution of this event cannot be individually traced, but one can compare the latitude band essentially devoid of vortices before the beginning of the episode, to the large quantity of vortices present later on (upper panel of Figure 14.5). Individual vortices in the image are 2000 km wide, and the average distance between adjacent vortices is 11000 ± 3500 km. From 2009, until the eruption of the 2010 GWS, storm activity in the storm alley was almost continuous, but there are very few high-resolution images of the region. Out of them, a few images taken at Saturn's darkest night at equinox (i.e. without the ring-shine illumination), are particularly relevant, since they imaged for the first time the presence of lightning on Saturn (Dyudina et al., 2010; section 14.3.2).

The evolution of the anticyclonic vortices created by storms can be traced in the images of Saturn captured by Cassini during its approach to Saturn, as can be seen in the animation PIA06083 (https://photojournal.jpl.nasa.gov/catalog/PIA06083). The anticyclones, between 2000 - 4000 km size east-to-west, were located at very close latitudes, and they moved at different speeds due to the meridional wind shear ($du/dy$), producing encounters and mergers. An example is shown in the lower panels of Figure 14.5, where it can be appreciated that during the mergers white clouds are generated around the anticyclones. These clouds, although not as bright as those of the storms, seem to have the same convective origin, as suggested by their morphology (Figure 5).

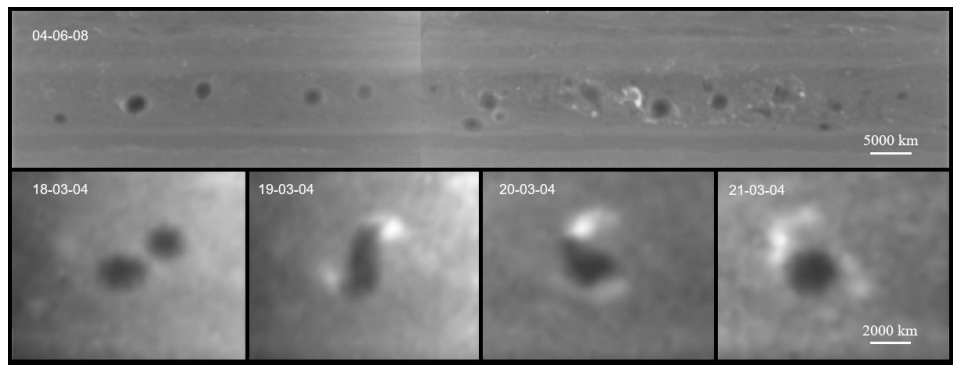

**Figure 14.5.** *Upper row: Planar projection of a section of the storm alley on 4 June 2008, with a large number of vortices and a storm visible. The projection covers 145° in longitude and 20° in latitude, centered at 35° S planetocentric latitude. Original images were taken with the ISS Wide Angle Camera and CB3 and IRP0 filters, at 15:03, 16:03 and 18:03. Lower row: A series of projections showing the merging of two vortices in March 2004. Projections cover 15° in longitude and 10° in latitude, centered at 35° S planetocentric latitude. Original images were taken with the Narrow Angle Camera and the CB2 and CL1 filters. Note that the scale, indicated with a white line, is different in the upper and lower row.*

### 14.2.3 Planetary-scale storms: Great White Spots

The rare and unique Great White Spots phenomenon in Saturn's atmosphere represents a major type of convective storm in the Solar System. The huge initial storm grows and expands in a few days forming a complex "head" that reaches a size greater than 10,000 km, which in its turn generates a turbulent wake that encircles the planet, i.e. a planetary-scale disturbance (Sánchez-Lavega, 1994; Sánchez-Lavega et al., 2018). So far, only six cases that follow this pattern have been observed, always in the northern hemisphere and with a periodicity around that of a Saturn year (see section 14.6). A complete list of all GWSs, together with the basic parameters that define them can be found in Table 13-1 in Sánchez-Lavega et al. (2018). Here we only give a global comparative view of all of the events according to their latitude.

(a) Equatorial phenomena

Three GWS events have been reported at the equator (years 1876, 1933, 1990) (Figures 14.6c, 14.12). The outbreaks occurred over a latitude range of 1.8°N to 9.8°N; the heads of the storms moved with the equatorial jet, close to the peak of the jet on its cyclonic side, with velocities ranging from $u$ = 365 to 400 ms$^{-1}$(see Fig. 14.14). These events are described in detail in Sánchez-Lavega (1982, 1994), and the GWS 1990 event and subsequent storms in Sánchez-Lavega et al. (1991, 1993, 1996), Beebe et al. (1992) and Barnet et al. (1992).

The equatorial region is abundant in spots, some of them white, although it is not clear if all of them correspond to convective phenomena. As an example, a large spot of size $\sim$ 27,000 km was observed in 1994-95 at 7.4°N moving with a velocity $u$ = 273 ms$^{-1}$. This spot did not follow the classic evolutionary pattern of a GWS, and it was probably a phenomenon related to the GWS 1990, in view of its proximity in time (Sánchez-Lavega et al., 1996). Also, in 2014 - 2015 a smaller bright white spot, some 7,000 km across, was observed at latitude 3.2°N moving at the peak of the jet with a velocity of $u$ = 445 ms$^{-1}$ (Sánchez-Lavega et al., 2016). In this case, the feature was formed by a cluster of spots and although its global temporal behavior differed from a global convective event, one cannot discard that each single spot in the cluster is convective in origin. However, Cassini was still in orbit in 2014 and 2015 and did not detect any SEDs.

(b) Mid-latitude events

Two GWS features have been reported at mid-latitudes. The first occurred in 1903, and there are no known photographs of this event. Reports at the time present drawings of the event and give transit times across the central meridian of the main spots forming the GWS (Sánchez-Lavega, 1982). Tracking of at least three spots is possible and gives an eastward velocity for the GWS of $u$ = 18 ms$^{-1}$. Determining the exact latitude of the main spot from drawings is imprecise, ranging between 31°N and 37°N. However, E. E. Barnard (1903) made precise measurements of the positions of the storm at the central meridian using a filar micrometer, which gave the latitude of its center at 36.7°N. In this latitude the zonal wind profile agrees with that of the GWS indicating the storm is on the anticyclonic side of the westward jet.

The second mid-latitude event is the best studied GWS so far. It took place in 2010-11 during the Cassini mission (Sánchez-Lavega et al., 2018 and references therein) at 32.4°N, moving westward with a velocity of -28 ms$^{-1}$ in the peak of the jet (Figure 14.6 d-f). The reader is referred to Sánchez-Lavega et al. (2018) for full details on the phenomenology related to this event: a rich and complex dynamics (vortices, waves and turbulence), vertical cloud structure and aerosol properties, chemical composition changes, temperature variability at different altitude levels and lightning events. It is interesting to note that these two GWS emerged in the same latitude band (westward jet) where the synoptic convective activity observed during the Voyager's epoch in 1980-81 took place.

(c) A polar GWS

In 1960, a GWS outbreak took place at latitude 52.5°N, close to a westward jet, and the head of the storm moved zonally with $u$ = 4 ms$^{-1}$. Most information about this storm comes from drawings (Dollfus, 1963), with only one low quality photograph of the phenomenon available (Figure 14.6 a-b). The available drawings show that after the storm developed zonally, expanding along the whole latitude band, with its clouds also expanding northwards until they reached the latitude of the hexagon (an unknown feature at the time) (Fig. 14.6b). The poleward expansion of the clouds took place with a velocity $v \sim 3$ ms$^{-1}$ (Sánchez-Lavega et al., 1994). This velocity is of the same order as that measured in the multiple storms in 2018 and 2020, which also stopped their northward progression at the latitude of the hexagon wave, as described in the previous section.

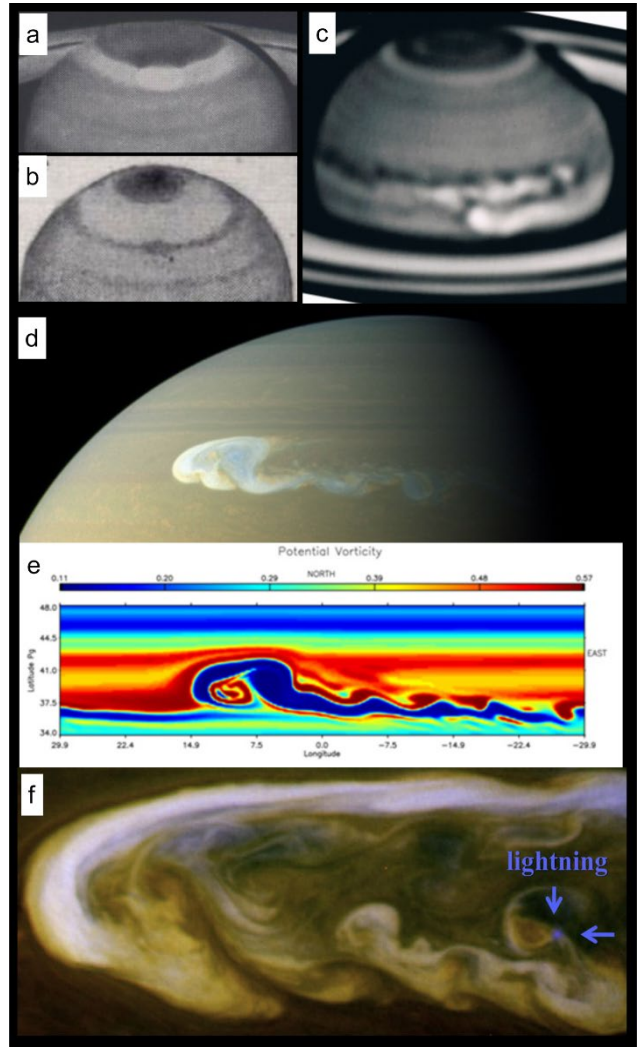

**Figure 14.6.** *A mosaic of selected Great White Spots. (a) and (b) 1960 GWS on 27 April and 2 August (drawings from Dollfus, 1963); (c) Mature stage of the 1990 GWS, on 5 November (Observatoire Pic-du-Midi, France); (d) GWS 2010 head and early tail on 24 December 2010 by Cassini ISS (image PIA1282), (e) EPIC simulation of the GWS 2010 12 days after the onset (from Sánchez-Lavega et al., 2011). Compare to (d); (f) The head of the GWS 2010 and a lightning event in an image taken by Cassini ISS on March 11, 2011 (from Dyudina et al., 2013).*

## 14.3 Cloud structure and lightning activity

### 14.3.1 Cloud properties

The structure and properties of the upper clouds forming the convective storms have been studied using radiative transfer analysis of two types of data. The first data set is based on sunlight reflected in the optical range using ground-based and HST images (from 225 nm to 954 nm, including the methane absorption bands at 727 nm and 890 nm). The second data set is based on the reflected and emitted radiation in the near infrared ($\sim 1$-5 μm) from Cassini Visual and Infrared Mapping Spectrometer (VIMS) observations.

Sromovsky et al. (2018) studied the 2008 storms and related dark ovals at 36.2°S. Based on a three-layer aerosol model, they found that the major differences between the background clouds, dark ovals and bright storms are due to the properties of an upper troposphere cloud (altitude level 0.2-0.4 bar) and a deep dense cloud layer (base at $\sim$ 3 bar). The tropospheric and deep clouds in the dark ovals have lower optical depths ($\tau$) than the background clouds. Storms are characterized by thick optical depth clouds of large $NH_3$-ice particles (radius ~ 7-9 μm) with $\tau$ (2 μm) ~ 5-18/bar. Above it, the tropospheric cloud has optical depths of 2-4 and is partly mixed with the lower cloud. Sromovsky et al. (2018) propose that during the active deep convection the ammonia cloud rises up to the level of the ubiquitous upper tropospheric layer without fully penetrating it, generating lightning (section 14.3.2). At this level, the divergence in the convective plume generates the dark anticyclonic oval, carrying part of the upper cloud layer outward where the large ammonia particles fall, leaving a thinned region in the upper cloud layer that looks darker at continuum wavelengths because it is less reflective.

The clouds in the 2018 polar storms were studied using HST images by Sánchez-Lavega et al. (2020). The tropospheric cloud in the storm had large optical depths $\tau$ ☐ 10 to 32 representing an increase in the particle density from ☐ 50 to 215 $cm^{-3}$ when compared to background clouds. In this model, the cloud top altitude reaches the ☐ 200 mbar level, elevated above the 600 mbar of the background clouds. The particles in the storm clouds are brighter and slightly larger (radius of 0.18 μm) relative to the background clouds.

In the case of the GWS 1990 and GWS 2010 the storm cloud tops, with optical depths $\tau$ ~ 15, reached the 300-400 mbar level, rising about 40 - 50 km relative to neighboring clouds at 1.4 bar (Acarreta and Sánchez-Lavega, 1999; García-Melendo et al., 2013). The tropospheric particles were found to be more reflectant at all wavelengths. The single scattering albedo at blue wavelengths was 0.99 at the storm head center compared to 0.90 in surrounding areas (Garcia-Melendo et al., 2013). This suggests that particles are coated by fresh $H_2O$-ice coming from deeper levels of the atmosphere (Sanz-Requena et al., 2012). From a near-infrared study of the GWS 2010, Sromovsky et al. (2013) propose a multi-component aerosol at the storm head, with fractions of 55% ammonia ice, 22% water ice, and 23% ammonium hydrosulfide. A firm conclusion from that study is that frozen water is indeed present in the convective storm head of the GWS 2010. This water ice is lofted ~200 km from above its freezing level near 10 bar as proposed by earlier models (Sánchez-Lavega and Battaner, 1987). Further details on the cloud structure of the GWS 2010 can be found in Sánchez-Lavega et al. (2018).

### 14.3.2 Lightning: Optical and radio emissions

The first indication of lightning in Saturn's atmosphere was obtained in November 1980 by the radio instrument on-board Voyager 1. Strong impulsive signals in the frequency range of a few MHz were detected and termed SEDs for Saturn Electrostatic Discharges (Warwick et al., 1981). Initially it was not clear if the SEDs came from discharges in the atmosphere or from the rings (Evans et al., 1982). Using an argument of visibility (the occultation by the planet was too long for a ring source), the ring hypothesis was refuted by Kaiser et al. (1983), and it was thought that the Voyager 1 SEDs stem from an atmospheric source in the equatorial region of Saturn. However, convective clouds were active at the time around a latitude of 35°N (Figure 14.1a), and it is possible that they were the source of the SEDs. Later on, combined radio and imaging observations by the Cassini mission have clearly established the atmospheric origin of Saturn's Electrostatic Discharges. These observations were made by the Cassini ISS (Porco et al., 2004) and the Cassini Radio Plasma Wave Science (RPWS) instrument (Gurnett et al., 2004).

The first direct link between SEDs and cloud features in Saturn's atmosphere was found when Cassini ISS imaged the so-called "dragon storm" in September 2004, and the RPWS instrument detected SEDs at the same time (Porco et al., 2005; Fischer et al., 2006). These combined observations revealed largely consistent longitudes and longitudinal drift rates of the storm and the SED source. Usually, SEDs can be detected by RPWS when the storm is on the side of Saturn facing Cassini, whereas they are practically absent when the storm is on the far side. However, due to an ionospheric propagation effect, some SEDs can also be detected when the storm is beyond the visible horizon (Zarka et al., 2006). The link between SEDs and cloud features was further strengthened when it was found that the white storm clouds were brighter when the SED rates were high (Dyudina et al., 2007; Fischer et al., 2007).

The first direct optical detection of Saturn lightning flashes was made in August 2009 around Saturn's equinox. During this time the reflected light from the rings towards the planet (ring-shine) was minimized, and so the Cassini cameras could spot flash-illuminated cloud tops with a diameter of about 200 km on Saturn's nightside (Dyudina et al., 2010). Figure 14.7 shows such bright spots in Cassini images from November 2009 in a large storm cloud with a size around 2000 km located at 36.2°S. The size of the bright spots suggested that the lightning came from 125-250 km below the cloud tops, which is where the water cloud layer is located in the atmosphere.

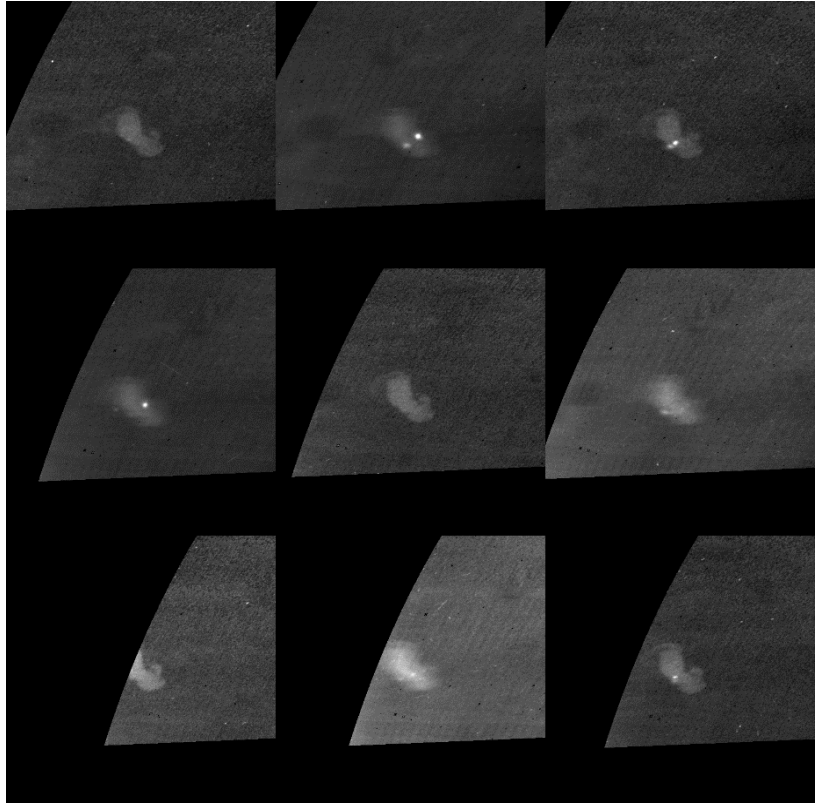

**Figure 14.7:** *Nine Cassini ISS images of Saturn's nightside from 30 November 2009 showing 200 km sized bright spots caused by lightning flashes (image PIA12576, ©NASA/JPL/SSI).*

On Earth, the charging of water cloud particles in thunderstorms is most effective in a temperature range of 248 K to 263 K (Rakov and Uman, 2003). On Saturn this temperature range is located at a level of 8-10 bars, about 200 km below the cloud tops, i.e. consistent with the altitude range found by Dyudina et al. (2010) within the water cloud layer. Another indication that the Saturn lightning source is in the water cloud layer comes from Cassini VIMS near-infrared spectra of the Great White Spot 2010. They revealed spectroscopic evidence for ammonia and water ices brought up to higher altitudes by strong vertical convection (Sromovsky et al., 2013). So, it is likely that the same particle charging mechanisms are at work on Saturn and Earth. As most of the sunlight is absorbed above the 2-bar level, Saturn's weather and thunderstorms at deep pressure levels must be powered by the planet's internal energy (Desch et al., 2006), which drives the vertical convection and brings up the water cloud to the visible atmospheric level, where it was observed as a bright eruption by Cassini ISS and VIMS. Dyudina et al. (2010) also measured the optical flash energy to be about $10^9$ J, which suggests that Saturn lightning is superbolt-like with total energies of about $10^{12-13}$ J (Turman, 1977; Fischer et al., 2011). Dyudina et al. (2013) could also identify visible flashes in blue wavelength (BL1 filter, 400-500 nm) on Saturn's dayside in the tail region of the GWS 2010 (Figure 14.6f). The optical flashes appeared in the cyclonic gaps between anticyclonic vortices where the cloud opacity was reduced, which allowed for the deep clouds to be more readily seen.

Figure 14.8 shows the number of detected SEDs per Saturn rotation as a function of time from the beginning of 2004 until the end of the Cassini mission in September 2017. The first SEDs were detected by Cassini RPWS in May 2004, and the last SEDs were seen in November 2013.

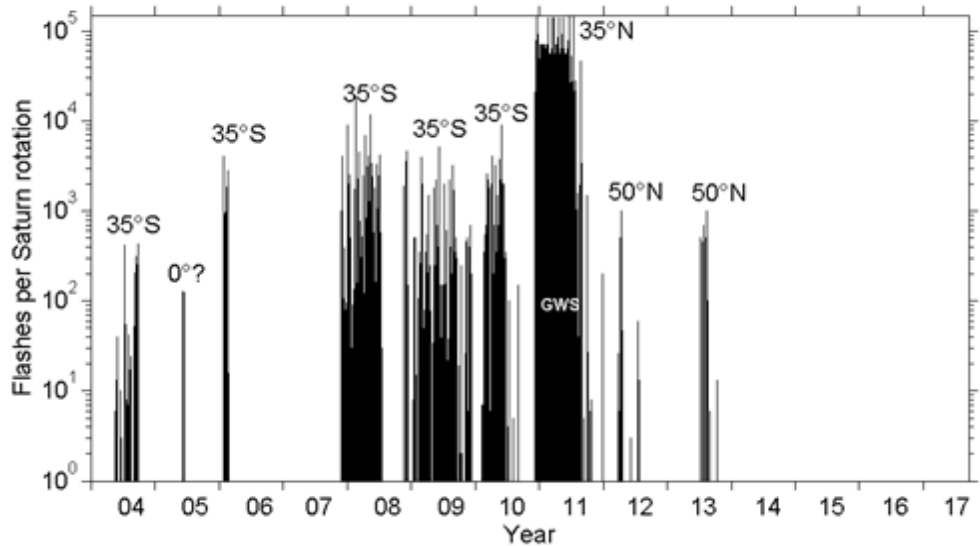

**Figure 8**: *Number of SEDs per Saturn rotation detected by Cassini RPWS as a function of time (years from 2004 to 2017). The planetocentric latitudes of the storms are indicated in the figure.*

Saturn lightning storms can last from a few days up to several months, and there was one storm that almost lasted throughout the year 2009. Most SEDs occurred □2 years around equinox (August 2009), and the SED storms switched from the southern hemisphere to the northern hemisphere one year after equinox, suggesting a seasonal influence (Fischer et al., 2011). There were also long-time intervals with no SED activity, especially from February 2006 until November 2007, and during the last 4 years of the mission. The absence of SEDs after November 2013 in the northern hemisphere could be explained by a kind of convective inhibition state of the atmosphere after the GWS of 2010/2011 (Li and Ingersoll, 2015). After July 2010 no more lightning storms occurred in the southern hemisphere, which might be due to the season turning into autumn there. The flash rate during the GWS was about 1-2 orders of magnitude higher than during the other synoptic-scale storms. For the GWS the flash rate was about 10 SEDs per second (Fischer et al., 2011), whereas the synoptic-scale storms typically had flash rates of a few SEDs per minute (Fischer et al., 2008).

During the Cassini mission, Saturn's thunderstorms only raged at specific latitudes, preferentially within the so-called "storm alleys" around 35°S (synoptic-scale storms) and 35°N (GWS 2010). At those latitudes there are broad minima in wind speed with small westward velocities with respect to Voyager's Saturn longitude system. The SED storms at 35°S usually consisted of just one convective cell at a time with one notable exception when two storm cells separated by 25° in longitude showed simultaneous SED activity for several months in spring 2008 (Fischer et al., 2019). The main storm cell of the GWS was located in its head, but several cells located in its growing tail were also sources of SEDs (Sayanagi et al., 2013). During the Cassini epoch there was potentially just one minor SED storm in the equatorial region which only lasted for about a week in June 2005 (Fischer et al., 2007). An interesting storm appeared at 50°N: A cyclonic vortex was already spotted by Cassini ISS in late 2006, but SEDs associated to it only appeared 4 years later in late 2010, around the same time when the GWS started to rage 15° further south. This vortex showed intermittent SED activity in 8 minor outbreaks until autumn 2013 (Gunnarson et al., 2023).

Table 14.1 shows a comparison between Saturn and terrestrial thunderstorms and flashes. There are few similarities, other than that both probably have similar charging mechanisms in water clouds around the freezing level. Storm sizes, durations, and frequencies are largely different between Saturn's lightning storms and terrestrial ones. On Saturn lightning storms are largely powered by internal heat of the planet (brought up to higher altitudes), whereas on Earth the energy finally comes from solar illumination. At both planets the atmospheres need to be in a condition to allow for the build-up of CAPE (Convective Available Potential Energy) and vertical moist convection. Flash and stroke durations have about the same order of magnitude, but the energy levels of Saturn's flashes are about 3 to 4 orders of magnitude higher. The strong radio power of SEDs led to their first detection with the large Earth-based radio telescope UTR-2 in Ukraine (Zakharenko et al., 2012; Konovalenko et al., 2013). In this way, and with the help of amateur astronomers, it is possible to continue investigating Saturn's lightning storms even after the end of the Cassini mission. For example, UTR-2 tried to detect SEDs during the polar storms in 2018 (Figure 14.1 c-d), but no detection was made in a few attempts. The spectral radio power of Saturn lightning only shows a weak fall-off ($1/f^{0.5}$) with increasing frequency, whereas Earth lightning falls off with $1/f^2$. The reason for this different spectral behavior in the frequency range of 1-40 MHz is unknown.

**Table 14.1:** Comparison between Saturn and Earth lightning storms. The values for the terrestrial storms were taken from the book of Rakov and Uman (2003)

| Characteristic | Saturn | Earth |
|---|---|---|
| Storm duration | days to several months | tens of minutes to hours |
| Frequency of storms | can be absent for years | ~1000 at any time |
| Storm size (regular) | 1500-3000 km | 25 km |
| Storm size (extraordinary) | 10 000 km in latitude (GWS) | up to 2000 km (hurricane) |
| Typical flash duration | 70 ms | 300 ms |
| Typical stroke duration | 100 μs | 30 μs |
| Size of illuminated cloud region (as viewed from above) | 200 km | 10 km |
| Flash optical energy | $10^{9-10}$ J | $10^6$ J |
| Flash total energy | $10^{12-13}$ J | $10^{8-9}$ J |
| Spectral radio power (at frequency f of a few MHz) | 10-100 W/Hz with $1/f^{0.5}$ | 0.1 W/Hz with $1/f^2$ |
| Pressure region | 8-10 bar (200 km below 1 bar) | 0.5 bar (at 5 km altitude) |
| Temperature range | around freezing level | -10°C to -25°C |
| Type of cloud | water cloud | water cloud |
| Storms powered by | internal heat of the planet | External heat from solar illumination |

**14.4 Moist convection in Saturn's atmosphere**

*14.4.1 Thermodynamics*

Lightning is intrinsically related to moist convection. In a hydrogen-dominated atmosphere, the effect of water vapor in a moist air parcel atmosphere is fundamentally different from that on Earth. Figure 14.9 compares the density of a saturated moist air parcel with the density of a dry air parcel from 320 K to 340 K at 16 bars on Saturn, which is approximately the level where water condenses. From left to right, as temperature increases, the density of a moist air parcel would decrease initially because density is inversely proportional to temperature. Then, the increase in temperature allows more water vapor in the air parcel and makes it heavier because water is heavier than hydrogen and helium. Therefore, passing a critical water mixing ratio, the density of a moist air parcel will increase with temperature. The two competing factors, temperature and the water abundance, equal at the critical water mixing ratio, which is:

$$q_v \ = \frac{R_v T}{(\epsilon - 1)(L_v - R_v T)} \tag{1}$$

where $\epsilon = \frac{m(H_2O)}{m(H_2/He)}$ is the molecular weight ratio between water and the ambient $H_2$-He mixture, $L_v$ is the latent heat of water, $R_v$ is the ideal gas constant of water vapor, and $T$ is the temperature at which water cloud forms (Li and Ingersoll, 2015). For Saturn, this value is around 0.9% as shown in Figure 14.9.

This non-monotonic behavior of air density with respective to temperature would inhibit convection as the upper layers of atmosphere cools, allowing convective available potential energy (CAPE, see section 14.4.2) to accumulate. Suppose that a planet has more than the critical value of water at the water condensation level and initially the moist air at the cloud condensation level has the same density as the air below the clouds. Radiative cooling to space moves the moist air parcel at the water condensation level from right to left along the green curve in Figure 14.9. Since the air parcel initially holds more than the critical value of water, its density will decrease due to condensation and precipitation, like dropping sandbags out of the hot air balloon. Convective motion is thus restricted to only the air above the cloud condensation level. During this stage, the moist air at the condensation level remains lighter than the air in the sub-cloud layer even though it is colder. This configuration is stable until the density of the moist air parcel starts to increase as temperature decreases (part of green curve to the left of the critical water mixing ratio). This state is called convection inhibition and it vanishes when the cold moist air parcel above the cloud has the same density as the warm moist air parcel below the cloud. Since Saturn has a thick atmosphere and a small radiative cooling flux, it takes decades for the shallow part of the atmosphere – above the cloud layer – to cool while suppressing deep convection. This process bears some resemblance to triggered convection on Earth, where CAPE is accumulated during favorable weather conditions and is released in thunderstorms. On Earth, some external dynamical trigger is required to force cloud parcels to break the layer of convective inhibition and convert the CAPE into vertical motion. Yet, the accumulation of CAPE and its subsequence release are all spontaneous on giant planets due to the thermodynamics of water in hydrogen atmosphere. The exact timing of convection, however, depends on the turbulent state of the atmosphere but the accumulation and release of CAPE are inevitable events as long as the water abundance exceeds the critical mixing ratio.

After convection is triggered, moist and warm air rises from deep layers, offsetting the dry and cold air above the cloud layer. As the mass ascends at the core of the convective region, the upper troposphere has a high pressure and is warm in temperature. A high-pressure center causes the warm air to expand, diverge and develop an anticyclonic circulation due to the rotation of the planet, similar to the mechanism of forming the anticyclonic circulation often seen near the tropopause in terrestrial hurricanes. The dynamic evolution of the radial-vertical stream-function and the concentration of the ammonia gas – a tracer constituent in Saturn's atmosphere that can be observed by Cassini RADAR -- have been simulated by a two-dimensional axisymmetric numerical model shown in Figure 14.10. First, the ammonia vapor is advected from the deep troposphere to the shallower layers and precipitates out. Second, the circulation reverses and advects ammonia-poor air back into the deep troposphere. Third, the kinetic energy of the radial-vertical motions is damped during the oscillation of the circulation. Finally, the averaged mixing ratio of ammonia is depleted with respect to a well-mixed state, in agreement with the 2.2 cm radar observation of the GWS 2010 (Janssen et al., 2013).

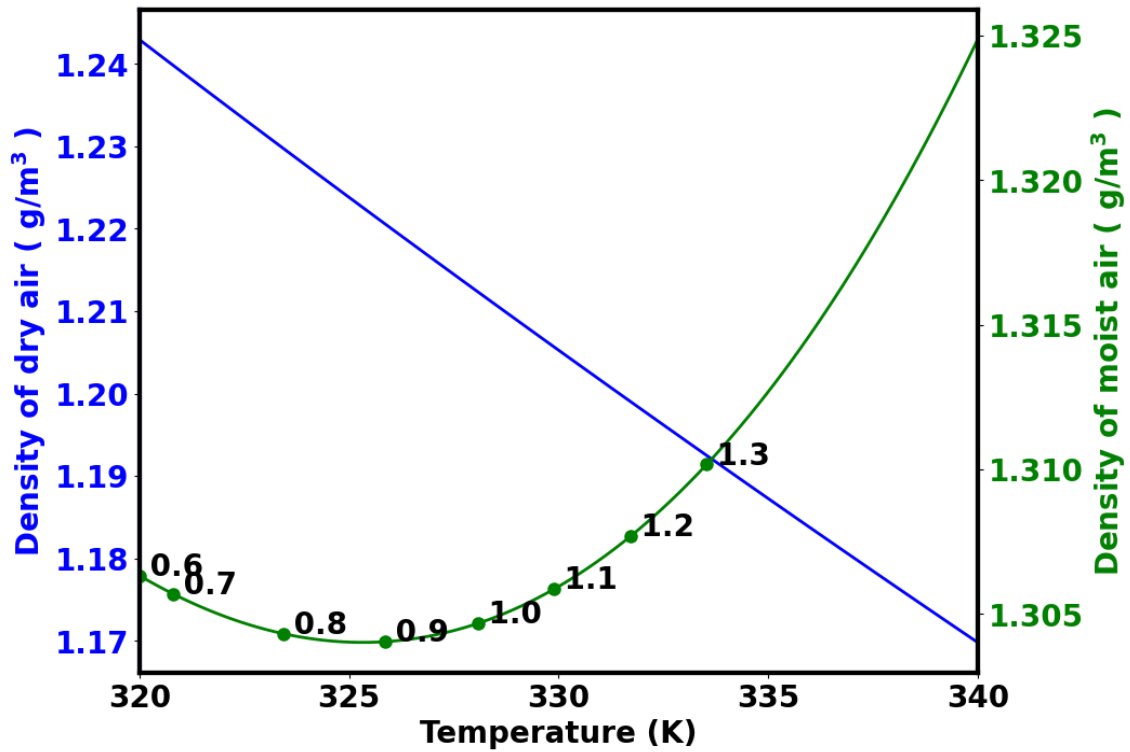

**Figure 14.9.** *Density of a saturated moist air parcel (green curve) compared to the density of a dry air parcel (blue line) at 16 bar pressure on Saturn. The water mixing ratio (%) is indicated in the figure on the green line. The minimum density of a moist air parcel is achieved at a water mixing ratio of around 0.9%.*

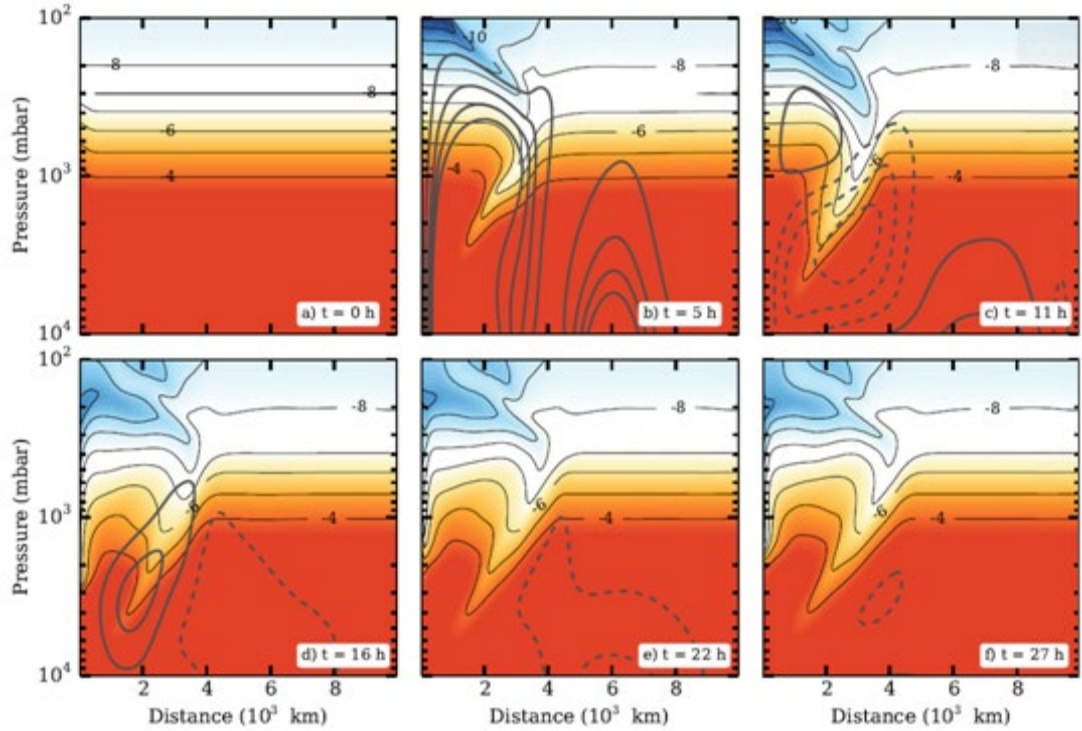

**Figure 14.10:** *Time evolution of ammonia vapor mixing ratio and the stream-function. Time is indicated at the bottom right corner increasing from panel (a) to panel (f). Colored shadowing and black contours show the mole mixing of ammonia vapor labeled by the exponent in base 10. Gray contours are the radial-vertical mass stream-functions; solid ones represent clockwise circulation and are drawn from $10^{12}$ kg/s to $7 \times 10^{12}$ kg/s at intervals of $2 \times 10^{12}$ kg/s; dashed ones represent counterclockwise circulation and are drawn from $-7 \times 10^{12}$ kg/s to $-10 \times 10^{12}$ kg/s at intervals of $2 \times 10^{12}$ kg/s (from Li and Ingersoll, 2015).*

### 14.4.2 One-dimensional convective models

The Convective Available Potential Energy (CAPE) at the ammonia and water cloud levels gives an estimate of the maximum velocities ($w$) expected for updrafts heated by the release of latent heat of condensation (Sánchez-Lavega, 2011)

$$CAPE = \frac{w_{max}^2}{2} = \int_{z_{LFC}}^{z_{LNB}} g\left(\frac{T'-T}{T}\right)dz \quad (\text{m}^2\text{s}^{-2} = \text{J kg}^{-1}) \tag{2}$$

where $Z_{LNB}$ is the level of neutral buoyancy and $Z_{LFC}$ is the level of free convection, T' and T are the temperatures of the parcel and the environment respectively, and $g$ is the acceleration of gravity. Then

$$w_{max} \approx \sqrt{2g\left(\frac{\Delta T}{<T>}\right)\Delta z} \tag{3}$$

The condensation of a volatile in a parcel would increase its temperature relative to its environment by

$$\Delta T = q_i \, \varepsilon \, \frac{L_i}{C_p} \qquad (4)$$

Here $q_i$ is the molar abundance (specific humidity) of water or ammonia, $\varepsilon = m_i/m$, where $m_i$ is the molecular weight of water (18.02 g mol$^{-1}$) or ammonia (17.03 g mol$^{-1}$) and $m = 2.14$ g mol$^{-1}$ is the mean molecular weight of the atmosphere, $L_i$ is the latent heat of condensation of water (2834x10$^3$ J kg$^{-1}$) or ammonia (1836x10$^3$ J kg$^{-1}$) and $C_p = 14{,}010$ J kg$^{-1}$K$^{-1}$ is the specific heat of Saturn's troposphere. For water clouds in Saturn, expected values are $\Delta T \sim 2\text{-}5$ K (for 1-3 times the solar abundance $q_{H2O} = 9.3x10^{-4}$), $<T> \sim 300$ K, $g \sim 9.5$ ms$^{-2}$, $\Delta z \sim 200$ km, and we find $w_{max} \sim 160\text{-}250$ ms$^{-1}$.

One-dimensional models of moist convection in the upper cloud layers of Saturn allow quantifying the initial conditions for the outbreak as well as the maximum height and ascent velocities (Sánchez-Lavega and Battaner, 1987). The momentum equation of the ascending parcel in a moist environment is given by (Sánchez-Lavega, 2011)

$$\frac{1}{2}\frac{dw'^2}{dz} = \underbrace{g\left(\frac{T_V{'} - T_V}{T_V}\right)}_{\text{Buoyancy Force}} + \underbrace{g\frac{\partial \pi_D}{\partial P}}_{\substack{\text{Dynamic} \\ \text{pressure}}} - \underbrace{C_D\frac{w'^2}{r_0}}_{\text{Friction Drag}} - \underbrace{\frac{1}{M}\frac{dM}{dz}w'^2}_{\text{Air entrainment}} - \underbrace{g\ell_C}_{\substack{\text{Weight} \\ \text{of condensable}}}$$

$$\qquad (5)$$

where $M$ is the vertical mass flux, $\ell_C$ is the condensate mass mixing ratio, $C_D$ is a drag coefficient, $r_0$ is the parcel size, $P$ is the pressure, $T_V$ is the virtual temperature that takes into account the humidity and vapor molecular weight, and $\pi_D$ the dynamical (perturbation) pressure that can be written as

$$\frac{\partial \pi_D}{\partial P} = \frac{1}{2}\frac{\partial \rho}{\partial P}w^2 + \frac{1}{2}\rho\frac{dw^2}{dP} \qquad (6)$$

The model depends on a variety of factors given in (5), which change with altitude, ambient relative humidity, deep water and ammonia abundances and the assumed $H_2$ ortho-para distribution. The entrainment rate is parameterized to the size of the entrained parcels as $(1/M)(dM/dz) \approx 0.2/r$, with $r$ being the entrained parcel size. Figure 14.11 shows resulting parcel vertical ascent velocities under representative values of the parameters in equation (5) (Sánchez-Lavega and Battaner, 1986). These models predict updrafts with maximum velocities in the most favorable cases between 150 and 200 ms$^{-1}$, and with cloud tops reaching altitudes between 150 to 300 mbar.

### 14.4.3 Three-dimensional mesoscale models

A 3D anelastic model with parameterized microphysics has been used to study the onset and evolution of water and ammonia moist convective storms up to sizes of $\sim 600$ km (Hueso and Sánchez-Lavega, 2004). The model tested different water concentrations (up to 5 times solar), ambient humidity in the range of 80-99%, molecular hydrogen (ortho-para distribution) and initial temperature disturbances. The model shows that water storms can be triggered by a thermal pulse in a wet (moist) ambient and be very energetic, reaching the 150 mbar level and developing vertical velocities on the order of 150 ms$^{-1}$, similar to those found by simple 1D models (Figure 14.11). Ammonia storms develop more easily than water storms, but with a much smaller intensity unless very large abundances of ammonia (10 times solar) are assumed to be present in Saturn's atmosphere. The presence of vertical wind shear and the Coriolis force outside the equator play a major role in shaping the morphology and properties of water-based storms. Simulations show that vertical columns of rising air due to water condensation have a horizontal morphology that depends on the intensity of convection, the latitude where they develop and the value of the vertical wind shear. The picture that emerges from the mesoscale simulations is that the head of synoptic and planetary scale storm systems (the GWS), can be formed by clusters of single cell thermals (300-600 km in size) mutually interacting in a complex way.

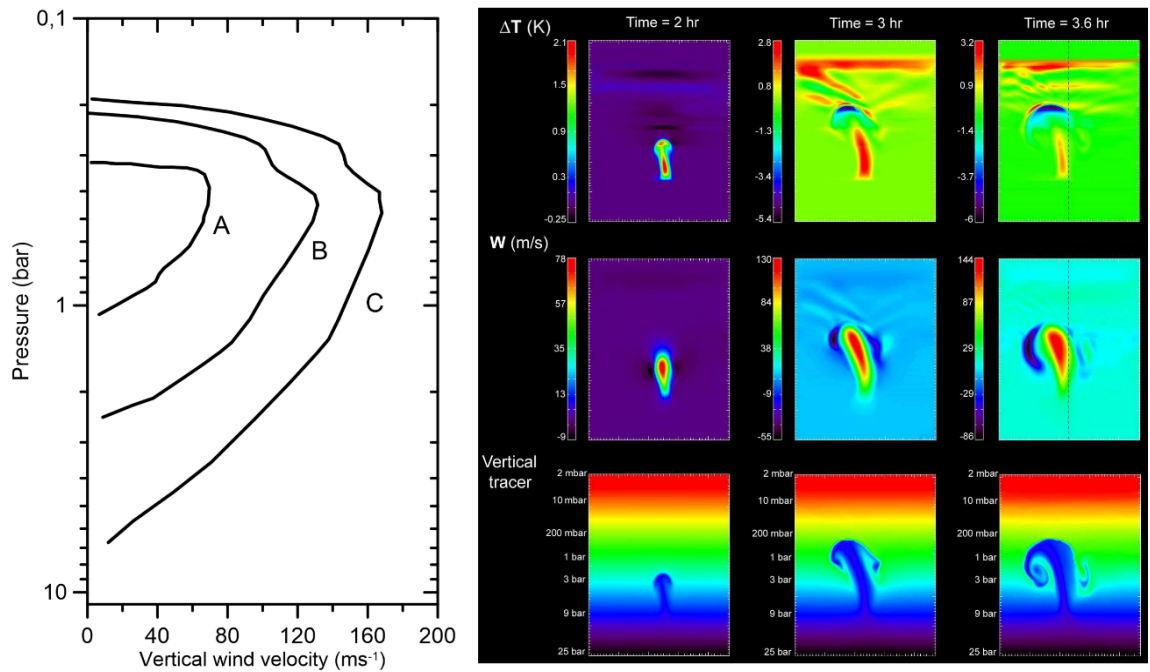

**Figure 14.11.** *Moist convection models in Saturn's atmosphere. Left panel: Results of a 1D parcel model for solar abundances of ammonia and water along a wet adiabat (relative humidity 100%), an intermediate distribution of $H_2$ and for different values of the entrainment rate of dry air in the parcel (r= 1, 10, 100 km for A , B, C, respectively). The dynamic pressure, drag force and weight of condensables are not considered in these cases. For a water solar abundance, saturation occurs at 12 bar (from Sánchez-Lavega and Battaner, 1986). Right panel: Water storm onset and development phases in a 3D simulation. The three columns show altitude-longitude cuts at the grid mid-plane of different storm variables at different times. The represented domain is 490 × 510 km. The first row shows storm temperature difference □T with the environment. The second row displays the corresponding vertical velocity (w), and the third row shows a passive tracer advected by the ascending plume. The time evolution goes from left to the right (from Hueso and Sánchez-Lavega,(2004).*

## 14.5 Numerical simulations of storms in Saturn's zonal winds

Synoptic and planetary-scale storms are large enough to be suited for investigation using simplified models of the atmosphere, where the weather layer can be considered very thin compared with its horizontal extension, such as for shallow water (SW) models (García-Melendo & Sánchez-Lavega, 2017) or for the EPIC General Circulation Model (Dowling et al., 1998). SW models have allowed exploration of the interacting dynamics between an established disturbance and the sheared flow in giant-planet tropospheres. Sugiyama et al. (2011, 2014) used a 2D convective cloud-resolving model with thermodynamics and microphysics ($NH_3$, $NH_4SH$, $H_2O$) to show, for the Jupiter case, the emergence of intermittent vigorous cumulonimbus clouds rising from the water condensation level to the tropopause. These kind of models have not yet been applied to Saturn clouds but similar results could be expected.

### 14.5.1 The SW and EPIC numerical models

In a one-layer or multilayer SW models, the atmosphere is modeled as a thin sheet of homogeneous fluid (Pedlosky, 1982) whereas in the EPIC model, the atmosphere is modeled as a hydrostatic multilayer model where the vertical coordinate is potential temperature. Both models solve the corresponding momentum equations for a moving fluid on a rotating ellipsoid. In the particular case of EPIC, the motion of fluid particles is considered adiabatic, so they are confined in horizontal layers in a stratified atmosphere. Shallow-Water and hydrostatic models such as EPIC cannot directly simulate convection. In fact, the SW model is decoupled from thermodynamics through the incompressibility condition ( $\nabla \cdot \vec{u} = 0$, where $\vec{u}$ is the velocity). Nevertheless, the energy and mass deposition in the weather layer associated to a convective source can be mimicked by introducing a perturbation. In the SW model, perturbations are introduced as local surface elevations on the free surface, whereas to perturb the EPIC model a source term is added to the Montgomery potential (Dowling et al., 1998). In most numerical models, these disturbances have a Gaussian shape in space and time, whose size and amplitude determine their intensity, but other geometries, such a top-hats, etc., are also implemented. In both models, grid resolution must be fine enough to resolve the perturbation with at least a few grid nodes. In most cases the perturbation is injected in the same way: every time step the precise longitude location of the source is computed according to its drift velocity; then the corresponding amplitude at the grid points inside the Gaussian source in the simulation domain are determined. In other studies (e.g. Sayanagi et al, 2007), the perturbation is introduced in the atmosphere with a drift speed which accommodates to the background zonal wind velocity. Numerical integration of the SW and primitive equations (e.g., as used in the EPIC model) allows the use of fully explicit schemes, which ensures parallelization by using a very efficient domain-decomposition strategy.

### 14.5.2 Simulation results

The large-scale dynamics of the Great Storm of 2010-2011 was successfully reproduced using the SW and EPIC models (Sánchez-Lavega et al., 2011; García-Melendo et al., 2013; García-Melendo & Sánchez-Lavega, 2017; Sánchez-Lavega et al., 2018, 2020, 2021). The SW experiments showed that the dynamics of planetary and intermediate scale storms is dominated by a combination of the interaction with the background zonal winds and the action of the Coriolis force. SW simulations of the GWS 1990, dominated by strong zonal wind shear and low latitude dynamics, suggest that the expansion of the storm in the equatorial region involves the generation of Rossby waves (Figure 14.12), a trapped gravity-Rossby wave at the equator, and probably Kelvin-Helmholtz instabilities produced by the advection of convective clouds in a zonal flow with high meridional shear (García-Melendo & Sánchez-Lavega, 2017).

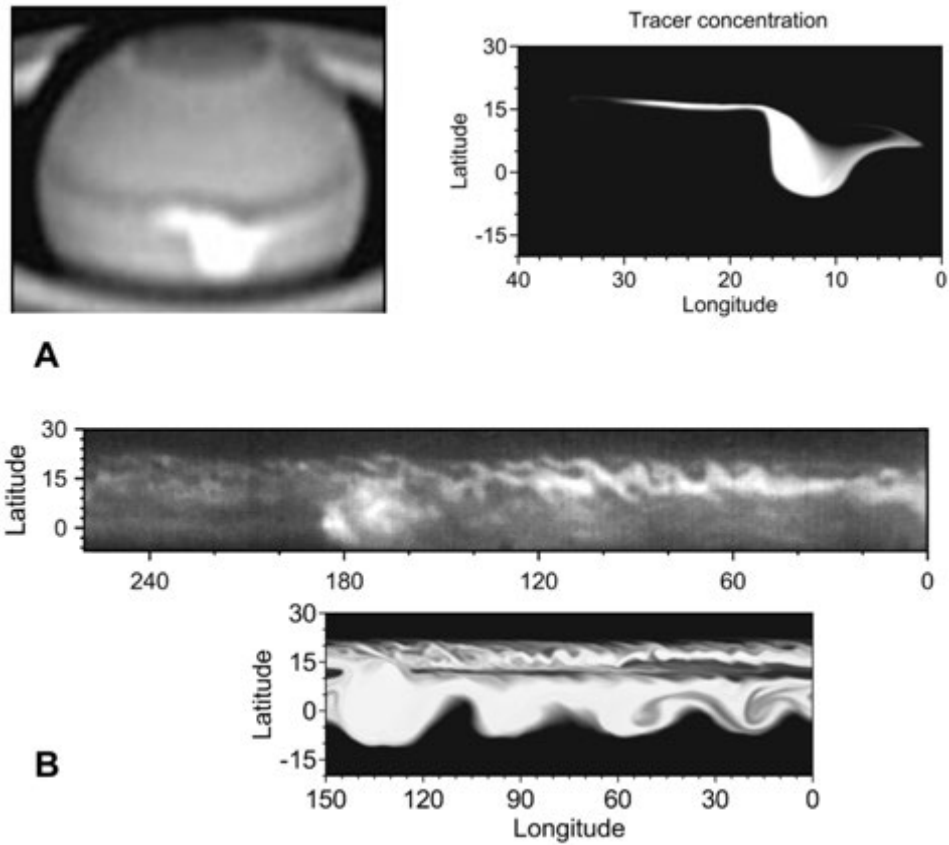

**Figure 14.12.** *(A) Left panel: GWS 1990 observed at Pic du Midi, 7 days after the onset on 2 October 1990 (Sánchez-Lavega et al., 1991). Right panel: 1-layer SW simulation of the GWS 1990 onset after 5 days (García-Melendo and Sánchez-Lavega, 2017). (B) Top panel: The GWS 1990 observed by HST on 17 November 1990, 53 days after the onset (Westphal et al., 1992) Bottom panel: 2-layer SW simulation after 40 days. Simulations represent a passive tracer whose concentration is coded between the arbitrary maximum values of 1.0 (white), to the black background (concentration = 0).*

In the 2010 GWS simulations, the divergence at the source of the perturbation adjusts to a geostrophic balance under the rapid rotation of Saturn to produce a strong anticyclonic circulation (García-Melendo et al., 2013; García-Melendo and Sánchez-Lavega, 2017). The anticyclonic cell is about ~ 10,000 km wide in latitude and moves westward 10 ms⁻¹ faster than the background wind (Figure 14.6e). It interacts with a cyclonic region on its northern side and with an anticyclonic region on the southern side generating a planetary scale disturbance that grows and propagates eastward of the source.

The SW and EPIC numerical simulations show that a combination of the source velocity, its intensity and its location on the zonal wind profile is what determines the width of the latitude band affected by the disturbance and its rapid zonal propagation, including vortex and wave generation. In the case of the 2010 GWS, the detailed reproduction of the dynamics during its first days also allowed sounding of the vertical structure of the zonal winds down to the 10 bar water cloud, suggesting that the winds must remain almost constant in the weather layer (Sánchez-Lavega et al., 2011).

High-latitude simulations of the 1960 GWS and the synoptic-scale 2018 and 2020 storms show that Coriolis forces play an important role there. The storms erupted in regions with different background zonal winds and shears and, in addition, the first 2018 storm erupted in the interior of a cyclonic vortex as indicated previously. The simulations were able to reproduce the size, motion, temporal evolution and eastward expansion constraining the source intensity of these storms (García-Melendo and Sánchez-Lavega, 2017; Sánchez-Lavega et al., 2020, 2021).

Modeling of the mid-latitude synoptic-scale storms with the SW model is peculiar since these storms do not consist of a persistent convective source. As discussed in section 14.2, convection in mid-latitudes was intermittent (although extended in time), concentrated in regions of a few thousand kilometers, and exhibited fast evolving morphology. SW simulations consisting of a small but strong source placed in the center of a larger but weaker disturbance within a turbulent background flow are able to reproduce the observed complex cloud patterns. The resulting patterns (see Fig. 14.13a) are a result of the strength and position of the inserted disturbances, advection by the imposed sheared zonal winds, and the background turbulence.

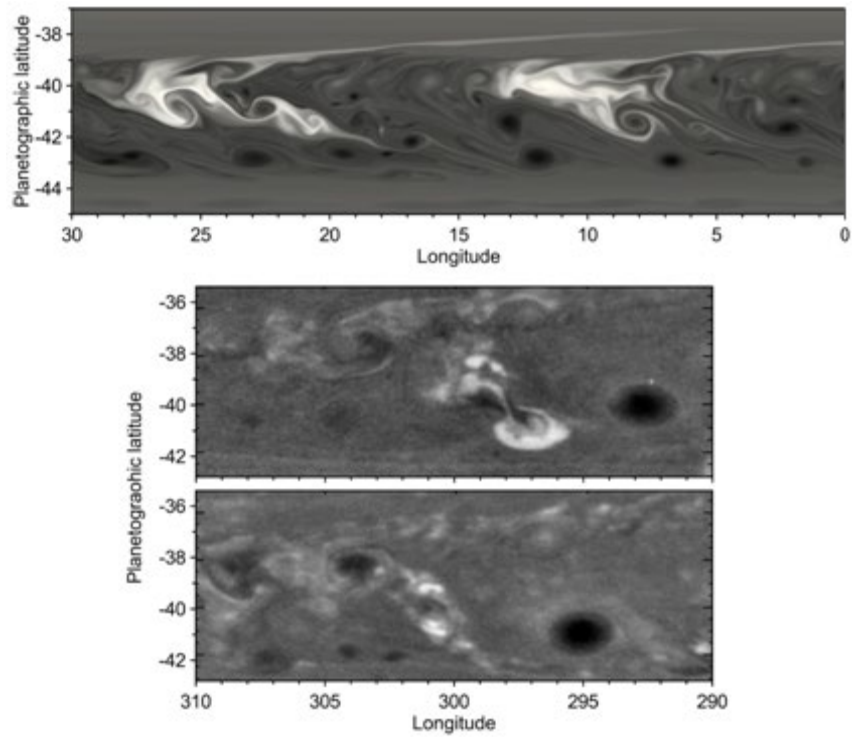

**Figure 14.13.** *The top panel represents SW simulations of the convective storms near the westward jet around 40°S. Storm clouds are represented by a passive tracer and superimposed on the background potential vorticity field seeded with turbulence. The two bottom panels are Cassini ISS observations of a convective storm observed in 2004. Cassini images are from Dyudina et al. (2007).*

One of the benefits of conducting the numerical simulations is that a comparison can be conducted of the intensity of the sources required to generate the different types of storms. To perform this comparison, models are run under the same numerical conditions since numerical diffusion depends on spatial resolution and time step. The SW simulations show that in terms of the injected mass flow volume ($m^3\ s^{-1}$), the 2018 and 2020 polar storms and the mid-latitude Cassini "storm-alley" storms require a significantly smaller mass flow, i.e., 0.01-0.1 of that required to generate a typical GWS (Sánchez-Lavega et al., 2020) ), to match the observed storm dynamics and morphologies.

## 14. 6 Discussion and conclusions

The number of convective events we present here is surely incomplete when considering the record of all reports of white spots on Saturn. Certainly, a number of other white spots compiled from visual observations in the 19th and 20th century and selected as plausible by Sánchez-Lavega (1982) could be convective events. Given the scarcity of information, they have not been included in this study. More recently, other white spots of synoptic-scale captured in ground-based photographic and digital images and with the HST, as described in Sánchez-Lavega et al. (1993, 1999; 2004a, 2004b) and Hueso et al. (2020), could possibly be convective events of less intensity and size than those presented here.

Figure 14.14 shows the distribution of the convective events against Saturn's zonal wind profile. It is striking that GWS features have only been observed in the northern hemisphere, perhaps due to a possible bias effect in the observation record of both hemispheres. There is a band of convective activity (latitudes ~ 1.5°N-10°N), centered on the maximum velocity of the equatorial jet, where three GWS cases have been reported. In the mid-latitudes (~ 35°N, ~ 35°S), the convective activity (synoptic-scale and GWS) is concentrated in two "storm alleys" in the westward jets in the northern and southern hemispheres. The northern subpolar region has also been active, with storms concentrated on both sides of a double jet, extending up to the latitude of the hexagon (75.5°N), while the 1960 GWS took place on a westward jet at 50°N where also a cyclonic vortex created intermittent SED activity from 2010 to 2013. In the subpolar southern hemisphere, there is only one reported storm placed again on the westward jet at 50°S. Therefore, it seems that the peak of the equatorial jet and the westward jets are the preferred places for convective activity. Del Genio et al. (2009) suggested that the preferred locations of convective activity in westward jets could be related to a combination of eddy momentum and energy fluxes with a meridional overturning circulation (see their Fig. 6.18).

Figure 14.15 shows the temporal distribution of the storms. As mentioned above most reported detections have been in the Northern Hemisphere. The GWS and polar storms outbreaks are concentrated in times close to the maximum of insolation at the top of the atmosphere, mainly between orbital longitudes ~ 90°-180°. The exception is the case of the 2010 GWS, ahead in time but at the same solar longitude and latitude as the storms observed in the Voyager epoch. Taken together, one can speculate with the existence of certain periodicities of approximately 30 and 60 years in these northern storms (Saturn's year is 29.4 Earth years) including (1) Period ~ 60 years for equatorial GWS in 1876-1933-1990; (2) Period ~ 30 years (annual cycle) between the mid-latitude storms in 1980(83)-2010, and (3) Period ~ 60 years for the polar storms in 1960 – 2018(20). Given the low statistics and the observational biases, these periodicities should be considered with caution. The partial visibility of the disk produced by the inclination of Saturn's axis of rotation and the occultation by the rings and their shadows could also introduce a bias in these periodicities. However, models of the interaction of moist convection of ammonia and water, and radiative cooling in hydrogen atmospheres produce oscillations that could lead to giant storm generation with a period of approximately 60 years as presented in section 14.4.1 (Li and Ingersoll, 2015). The results of the 2D convective model by Sugiyama et al. (2014), show that the storm intermittency period is controlled by the radiative cooling of the atmosphere and is approximately obtained by dividing the mean temperature increase due to the release of latent heat in the storm's cumulonimbus clouds and the radiative cooling rate. The radiative time constant is of the order of a Saturn year in its upper troposphere (Conrath et al., 1990).

The observational record suggests a storm deficit in the southern hemisphere. The observations of that hemisphere by Cassini and from the ground and HST reveal however a long period of synoptic-scale storms that lasted about 8 years (2002 - 2010). This activity occurred during the period of maximum insolation in that hemisphere. The southern 1994 storm could be considered, within an observational sense, to be an anomalous event because of the low insolation at this latitude. Probably an observational bias in the hemispheric visibility is playing a role in this "GWS southern void". We note that the hemispherical asymmetry in insolation (Figure 14.15) with larger values in the southern hemisphere is due to Saturn's elliptical orbit around the Sun in which perihelion occurs near southern summer solstice.

Mid-latitude storms (latitudes ~ 35°) generate dark anticyclones as observed in the storm alley in the north (Sromovsky et al., 1983), in detail in the south (Dyudina et al., 2007; Baines et al., 2009; Sromovsky et al., 2018; Fischer et al., 2019), and for the 2010 GWS (Sánchez-Lavega et al., 2012, Sayanagi et al., 2013, Hueso et al., 2020). Two examples of major anticyclones in these latitudes are BS1 (informally called Brown Spot 1) that generated from the convective activity in the Voyager epoch in 1980-81 (Smith et al., 1981, 1982; García-Melendo et al., 2007) and the AV (Anticyclone Vortex) formed in the wake of the GWS 2010 storm. BS1 had a size of $6,100 – 4,500$ km (east-west and north-south) and survived at least for 1 year (García-Melendo et al., 2007). AV has survived at least 10 years, shrinking from an initial size of 11,000 - 12,000 km to its current size of 5,700 - 3,000 km. The relative vorticity measured in these anticyclones was $-(4.0\pm1.5)\text{x}10^{-5}$ $\text{s}^{-1}$ for BS1 and $-(6.0\pm1.0)\text{x}10^{-5}$ $\text{s}^{-1}$ for AV, above that of the background meridional shear of $du/dy \sim -3.0\text{x}10^{5}$ $\text{s}^{-1}$, but much lower than the Coriolis vorticity $f = 2.2\text{x}10^{-4}$ $\text{s}^{-1}$. Although the SW and EPIC numerical simulations show vortex formation in the wake of the storm sources (Sánchez-Lavega et al., 2012; García-Melendo and Sánchez-Lavega, 2017), a proper understanding of how these anticyclones form from the energy carried by the convective outbreak within a sheared flow under the influence of planetary vorticity and including full thermodynamics is still lacking. All this raises the question of whether other anticyclones on Saturn have had their origin in convective storms, as for example the large (11,000 - 3,700 km) and long-lived North Polar Spot (NPS) observed between 1980-81 and 1996 at 71.7°N (Sánchez-Lavega et al., 1997). If this is the case, the presence of closed anticyclonic ovals on Saturn would be a sign of a previous convective storm.

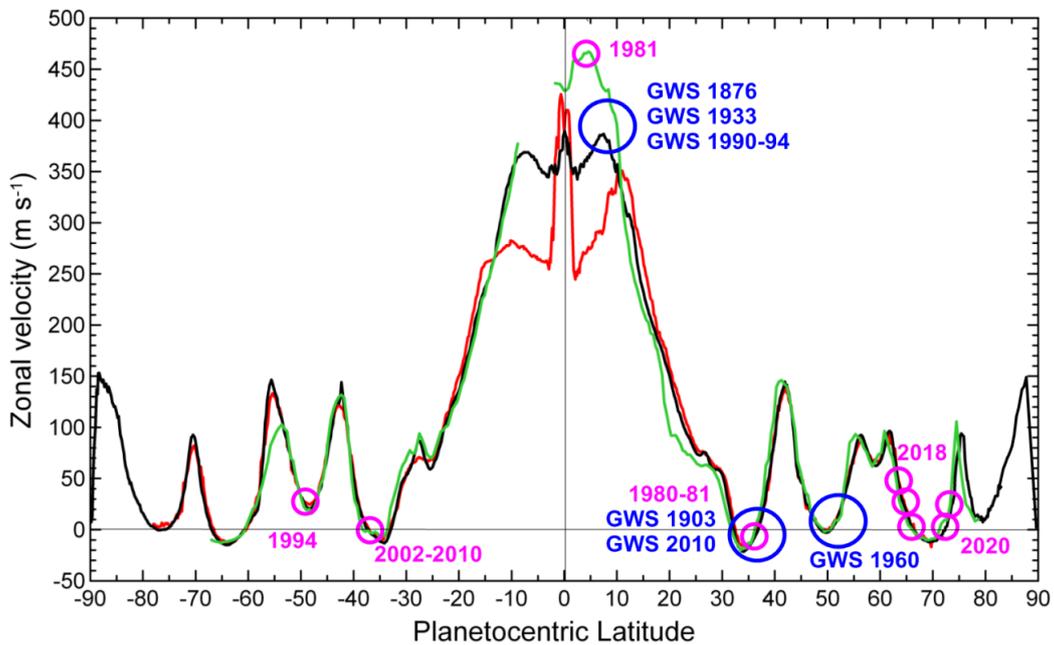

**Figure 14.14.** *Saturn's zonal wind profile at cloud level and observed convective events. The green line is from Voyager 1 and 2 in 1980-81, the black line is from Cassini ISS in 2004-2009 at wavelengths sensing motions of low altitude clouds, and the red line is from Cassini ISS at wavelengths sensing motions at high altitude levels (green from Sánchez-Lavega et al., 2000; black and red from García-Melendo et al., 2011 with north polar jet profile from Antuñano et al., 2018). Blue circles mark the locations of the GWS major events and magenta circles mark the locations of synoptic-scale storms. Intermittent SEDs events were detected from 2010-13 at 50°N, not shown because no optical related features were reported.*

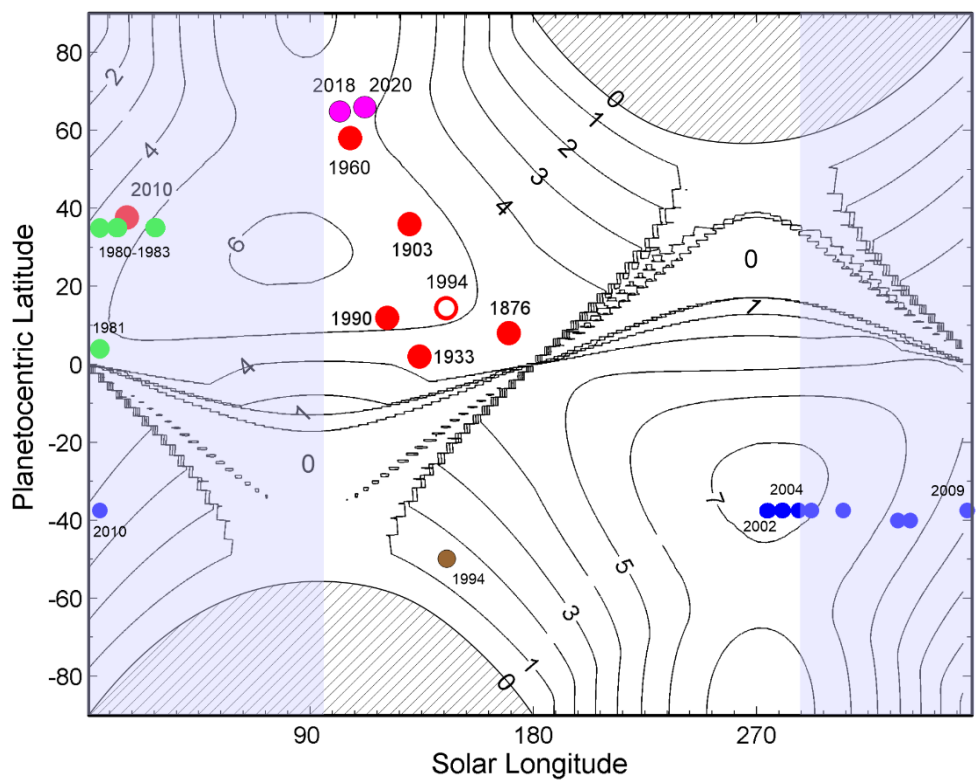

**Figure 14.15.** *Daily insolation at the top of the atmosphere of Saturn along an orbital cycle and major known convective events. The insolation curves are given in Wm⁻² and the solar longitude locates the planet position in its orbit. The blue shaded area is the period covered by the Cassini mission. The convective events are identified by the color dots, red for the GWS cases and the rest is for the synoptic-scale events described in the text. Adapted from Sánchez-Lavega et al., 2018).*

Numerous aspects remain open about our understanding of the mechanisms subjacent to convective storm formation in the atmospheres of the giant planets, with very similar clouds and thermodynamics. In the case of Saturn, we need to know the spatial and temporal distribution of water vapor below the upper aerosol layers since it appears as the main agent of moist convection. We also need to know about the role played in storm triggering by the seasonal insolation cycle as modulated by the tilt and flattening of the planet and by the shadow and radiation from the rings. Future models will need to investigate how the general circulation of the atmosphere, i.e., the zonal wind profile and the possible existence of meridional circulation cells in the upper troposphere influence storm formation. Such modeling efforts may confirm whether the existence of temporal cycles are involved in the development of synoptic and planetary scale storms (e.g., GWS).

No space mission is planned for the Saturnian system in the coming years, with the exception of Dragonfly scheduled to arrive at Titan in 2034. Observational studies of Saturn will have to rely on space (HST, JWST) and ground-based telescopes. The survey of storm activity over time and other phenomena in Saturn's atmosphere will be the domain of the small telescopes employed mainly by amateur astronomers, who provide nearly continuous coverage of the planet. The next two ring-plane crossings will take place on March 23, 2025 (for the Earth) and May 6, 2025 (for the Sun), when both hemispheres will be fully visible simultaneously. Thereafter, the southern hemisphere will be visible for a long period of time and in May 12, 2032 the south pole of Saturn will be most inclined toward Earth. Then, the next ring crossings will occur between October 15, 2038 and July 9, 2039. We therefore have a long period of more than 20 years ahead of us to see if the "great southern storm void" in GWS activity is a real phenomena, or, if not, detect such an outbreak for the first time in the southern hemisphere.

**References**

Acarreta, J. R., and A. Sánchez-Lavega (1999). Vertical cloud structure in Saturn's 1990 Equatorial storm. *Icarus* 137, 24–33.

Antuñano A., T. del Río-Gaztelurrutia, A. Sánchez-Lavega, and J. Rodriguez Aseguinolaza (2018). Cloud Morphology and Dynamics in Saturn's Northern Polar Region. *Icaru*s 299, 117-132.

Archinal, B. A. (list next 2 authors, then et.al) et al. (2018) Report of the IAU working group on cartographic coordinates and rotation elements: 2015. *Celest. Mech. Dyn. Astr.* 130, 22. https://link.springer.com/article/10.1007/s10569-017-9805-5

Atreya, S. K., M. H. Wong, T. C. Owen, et al;. (1999). A Comparison of the Atmospheres of Jupiter and Saturn: Deep Atmospheric Composition, Cloud Structure, Vertical Mixing, and Origin. *Planet. Space Sci.* 47, 1243-1262.

Baines K. H., M. L. Delitsky, T. W. Momary, et al. (2009). Storm clouds on Saturn: Lightning-induced chemistry and associated materials consistent with Cassini/VIMS spectra. *Planet. Space Sci.* 57, 1650-1658,   doi:10.1016/j.pss.2009.06.025.

Barnard, E. E. (1903). White Spot on Saturn, *Astron. J.* 23, 143-144.

Barnet, C. D., J. A. Westphal, R. F. Beebe, and L. F. Huber (1992). Hubble space telescope observations of the 1990 equatorial disturbance on Saturn: Zonal winds and central Meridian albedos. *Icarus* 100, 499-511. Doi: 10.1016/0019-1035(92)90113-L.

Beebe, R. F., C. Barnet, C., P. V. Sada, and A. S. Murrell (1992). The onset and growth of the 1990 equatorial disturbance on Saturn. *Icarus* 95, 163-172. Doi: 10.1016/0019-1035(92)90035-6.

Conrath B. J., P. J. Gierasch, and S. S. Leroy (1990). Temperature and circulation in the stratosphere of the outer planets. *Icarus* 83, 255 – 281.

Del Genio A. D., R. K. Achterberg, K. H. Baines, et al. (2009). Saturn Atmospheric Structure and Dynamics, Chapter 6 in Saturn after Cassini-Huygens. M. Dougherty, L. Esposito and T. Krimigis (edts.), Springer-Verlag, pp. 113-159.

Desch, M. D., and L. M. Kaiser (1981). Voyager measurements of the rotation period of Saturn's magnetic field. *Geophys. Res. Lett.*, 8, 253–256.

Desch, M. D., G. Fischer, M. L. Kaiser, W. M. Farrell, W. S. Kurth, D. A. Gurnett, P. Zarka,  A. Lecacheux, C. C. Porco, A. P. Ingersoll, and U. A. Dyudina (2006). Cassini RPWS and Imaging Observations of Saturn Lightning, in Planetary Radio Emissions VI, edited by Rucker, H. O., Kurth, W. S., and Mann, G., Austrian Academy of Sciences Press, Vienna, 103-110.

Dollfus, A. (1963). Mouvements dans l'Atmosphère de Saturne en 1960. Observations coordonées par l'Union Astronomique Internationale. *Icarus*, 2, 109-114.

Dowling, T. E., A. S. Fischer, P. J. Gierasch, et al. (1998). The Explicit Planetary Isentropic-Coordinate (EPIC) Atmospheric Model.  *Icarus* 132, 221-238. Doi: 10.1006/icar.1998.5917.

Dyudina, U. A. , A. P. Ingersoll, S. P. Ewald, et al. (2007). Lightning Storms on Saturn Observed by Cassini ISS and RPWS During 2004-2006, *Icarus*, 190, 545-555, doi:10.1016/j.icarus.2007.03.035.

Dyudina, U. A. , A. P. Ingersoll, S. P. Ewald, et al. (2010). Detection of Visible Lightning on Saturn, *Geophys. Res. Lett.*, 37, L09205, doi:10.1029/2010GL043188.

Dyudina, U. A. , A. P. Ingersoll, S. P. Ewald, et al. (2013). Saturn's Visible Lightning, Its Radio Emissions, and the Structure of the 2009-2011 Lightning Storms, *Icarus*, 225, 1029-1037, doi:10.1016.j.icarus.2013.07.013.

Evans, D. R., J. H. Romig, C. W. Hord, et al. (1982). The source of Saturn electrostatic discharges, *Nature*, 299, 236-237, doi:10.1038/299236a0.

Fischer, G., M. D Desch, P. Zarka, et al. (2006). Saturn Lightning Recorded by Cassini/RPWS in 2004, *Icarus*, 183, 135-152, doi:10.1016/j.icarus.2006.02.010.

Fischer, G., W. S. Kurth, U. A. Dyudina, et al. (2007). Analysis of a Giant Lightning Storm on Saturn, *Icarus*, 190, 528-544, doi:10.1016/j.icarus.2007.04.002.

Fischer, G., D. A. Gurnett, W. S. Kurth, et al. (2008). Atmospheric Electricity at Saturn, *Space Sci. Rev.*, 137, 271-285, doi:10.1007/s11214-008-9370-z.

Fischer, G., W. S. Kurth, D. A. Gurnett, et al. (2011). A Giant Thunderstorms on Saturn, *Nature*, 475, 75-77, doi:10.1038/nature10205.

Fischer, G., J. A. Pagaran, P. Zarka, et al. (2019). Analysis of a long-lived, two-cell lightning storm on Saturn, *Astron. Astrophys.*, 621, id.A113, 19 pp, doi: 10.1051/0004-6361/201833014.

García-Melendo E., A. Sánchez-Lavega, and R. Hueso (2007). Numerical models of Saturn's long-lived anticyclones, *Icarus* 191, 665-677.

García-Melendo, E., S. Pérez-Hoyos, A. Sánchez-Lavega, et al. (2011). Saturn's zonal wind profile in 2004–2009 from Cassini ISS images and its long-term variability. *Icarus* 215, 62–74.

García-Melendo E., R. Hueso, A. Sánchez-Lavega, et al. (2013). Atmospheric dynamics of Saturn's 2010 giant storm. *Nature Geoscience* 6, 525–529.

García-Melendo, E. and A. Sánchez-Lavega (2017). Shallow water simulations of Saturn's giant storms at different latitudes. *Icarus* 286, 241–260.

Gunnarson, J., K. M. Sayanagi, G. Fischer et al. (2022). Multiple convective storms within a single cyclone on Saturn. *Icarus*, in review.

Gurnett, D. A. . (Add two more authors, then:)et al. (2004). The Cassini Radio and Plasma Wave Science Investigation. *Space Sci. Rev.*, 114, 395-463, doi:10.1007/s11214-004-1434-0.

Hueso, R., and A. Sánchez-Lavega (2004). A three – Dimensional Model of Moist Convection for the Giant Planets II: Saturn's water and ammonia moist convective storms. *Icarus*, 172, 255-271.

Hueso, R., A. Sanchez-Lavega, J. F. Rojas et al. (2020). Saturn atmospheric dynamics one year after Cassini: Long-lived features and time variations in the drift of the Hexagon. *Icarus* 336, 113429. Doi: 10.1016/j.icarus.2019.113429.

Hunt, G. E., D. Godfrey, J.-P. Muller, et al. (1982). Dynamical features in the northern hemisphere of Saturn from Voyager 1 images. *Nature*, 297, 132-134.

Ingersoll, A. P., R. F. Beebe, B. J. Conrath et al. (1984). Structure and dynamics of Saturn's atmosphere, in Saturn, Gehrels, T., and M. S. Matthews, University of Arizona Press, 195–238.

Ingersoll A. P. (2020). Cassini exploration of the Planet Saturn: A Comprehensive Review. *Space Sci. Rev.*, 216: 122.

Janssen, M. A., A. P. Ingersoll, M. D. Allison, et al. (2013). Saturn's thermal emission at 2.2-cm wavelength as imaged by the Cassini RADAR radiometer. *Icarus* 226, 522-535.

Kaiser, M.L. , J.E.P. Connerney, and M. D. Desch (1983). Atmospheric storm explanation of Saturnian electrostatic discharges. *Nature*, 303, 50-53, doi:10.1038/303050a0.

Konovalenko, A. A., N. N. Kalinichenko, H. O. Rucker et al. (2013). Earliest recorded ground-based decameter wavelength observations of Saturn's lightning during the giant E-storm detected by Cassini spacecraft in early 2006. *Icarus* 224, 1, 14-23.

Li, C., and A. P. Ingersoll (2015). Moist convection in hydrogen atmospheres and the frequency of Saturn's giant storms. *Nature Geoscience*, 8, 398-403, doi:10.1038/ngeo2405.

Mendikoa I., A. Sánchez-Lavega, S. Pérez-Hoyos, et al. (2016). PlanetCam UPV/EHU: A two channel lucky imaging camera for Solar System studies in the spectral range 0.38-1.7 μm. *Pub. Astron. Soc. Pacific* 128, 035002 (22 pp).

Pedlosky, J. (1982). Geophysical fluid dynamics. Springer-Verlag.

Porco, C. C., R. A. West, S. Squyres et al. (2004). Cassini Imaging Science: Instrument Characteristics and Anticipated Scientific Investigations at Saturn. *Space Sci. Rev.*, 115, 363-497, doi:10.1007/s11214-004-1456-7.

Porco, C. C., E. Baker, J. Barbara, et al. (2005). Cassini Imaging Science: Initial Results on Saturn's Atmosphere. *Science*, 307, 1243-1247, doi:10.1126/science.1107691.

Rakov, V. A, and M. A. Uman (2003). Lightning, Cambridge Univ. Press, Cambridge, UK.

Sánchez Lavega A. (1982). Motions in Saturn's Atmosphere: Observations before Voyager Encounters. *Icarus* 49, 1 - 16.

Sánchez-Lavega A. and E. Battaner (1986). Long-term changes in Saturn's atmospheric belts and zones. *Aston. Astrophys. Supp. Ser.* 64, 287-301.

Sánchez-Lavega A. and E. Battaner (1987). The nature of Saturn's Great White Spots. *Astron. Astrophys.* 185, 315-326.

Sánchez-Lavega, A., F. Colas, J. Lecacheux, et al. (1991). The Great White Spot and disturbances in Saturn's equatorial atmosphere during 1990. *Nature* 353, 397-401. Doi: 10.1038/353397a0.

Sánchez-Lavega A., J. Lecacheux, F. Colas, et al. (1993). Temporal behavior of cloud morphologies and motions in Saturn's atmosphere. *J. Geophysical Research* 98, 18857 - 18872.

Sánchez Lavega, A., J. Lecacheux, J. M. Gomez, et al. (1996). Large-scale storms in Saturn's atmosphere during 1994. *Science* 271, 631–634.

Sánchez Lavega A., J. F. Rojas, J. R. Acarreta, et al. (1997). New Observations and studies of Saturn's long-lived North Polar Spot. *Icarus*, 128, 322-334.

Sánchez Lavega A., J. Lecacheux, F. Colas, et al. (1999). Discrete cloud activity in Saturn's equator during 1995, 1996 and 1997. *Planetary Space Sciences*, 47, 1277 – 1283.

Sánchez-Lavega, A., J. F. Rojas and P. V. Sada (2000). Saturn's zonal winds at cloud level. *Icarus*, 147, 405–420.

Sánchez-Lavega, A., S. Pérez-Hoyos, J. F. Rojas et al. (2003) A strong decrease in Saturn's equatorial jet at cloud level, Nature, 423, 623-625.

Sánchez-Lavega, A., R. Hueso, S. Pérez-Hoyos et al. (2004a). Saturn's Cloud Morphology and Zonal Winds Before the Cassini Encounter. *Icarus*, 170, 519-523.

Sánchez-Lavega A., S. Pérez-Hoyos, and R. Hueso (2004b). Condensate clouds in planetary atmospheres: a useful application of the Clausius-Clapeyron equation. *Amer. J. Physics* 72, 767-774.

Sánchez-Lavega, A. (2011). An Introduction to Planetary Atmospheres, Taylor-Francis, CRC Press, Florida, pp. 629.

Sánchez-Lavega A., T. del Río-Gaztelurrutia, R. Hueso, et al. (2011). Deep winds beneath Saturn's upper clouds from a seasonal long-lived planetary-scale storm. *Nature* 475, 71-74. Doi: 10.1038/ nature10203.

Sánchez-Lavega A., T. del Rio-Gaztelurrutia, M. Delcroix et al. (2012). Ground-based observations of the long-term evolution and death of Saturn's 2010 Great White Spot. *Icarus* 220, 561-576.

Sánchez-Lavega A., E. García-Melendo, S. Pérez-Hoyos et al. (2016). An Enduring rapidly moving storm as a guide to Saturn's equatorial jet complex structure. *Nature Communications* 7, 13262.
DOI 10.1038/NCOMMS13262.

Sánchez-Lavega A., G. Fischer, L. N. Fletcher, E. Garcia-Melendo, B. Hesman, S. Perez-Hoyos, K. Sayanagi and L. Sromovsky (2018). The Great Storm of 2010-2011, Chapter 13 of the book Saturn in the 21st Century, pp. 377-416, eds. K. H. Baines, F. M. Flasar, N. Krupp, T. S. Stallard, Cambridge University Press (Cambridge, U.K.)

Sánchez-Lavega A., L. Sromovsky, A. Showman, A. Del Genio, R. Young, R. Hueso, E. García Melendo, Y. Kaspi, G. S. Orton, N. Barrado-Izagirre, D. Choi, J. Barbara, *"Zonal Jets in Gas Giants"* chapter of the book *Zonal Jets*, pp. 9-45, eds. B. Galperin and P. Read, Cambridge University Press (2019). ISBN 978-1-107-04388-6 Hardback.

Sánchez-Lavega A., E. García-Melendo, J. Legarreta, et al. (2020). A complex storm system in Saturn's north polar atmosphere in 2018. *Nature Astronomy* 4, 180-187. Doi: 10.1038/s41550-019-0914-9.

Sánchez-Lavega A., E. García-Melendo, R. Hueso, et al. (2021). Interaction of Saturn's Hexagon with convective storms. *Geophysical Research Letters,* 48, e2021GL092461. https://doi.org/10.1029/2021GL092461

Sanz-Requena J. F., S. Perez-Hoyos, A. Sanchez-Lavega, et al. (2012). Cloud structure of Saturn's 2010 storm from ground-based visual imaging. *Icarus* 219, 142-149.

Sayanagi K. M., U. A. Dyudina, S. P. Ewald, et al. (2013). Dynamics of Saturn's great storm of 2010–2011 from Cassini ISS and RPWS. *Icarus* 223, 460–478. Doi: 10.1016/j.icarus. 2012.12.013.

Smith B. A., L. Soderblom, R. F. Beebe et al. (1981). Encounter with Saturn: Voyager 1 Imaging Science Results. *Science* 212, 163-191. Doi: 10.1126/science.212.4491.163.

Smith B. A., L. Spderblom, R. Batson et al. (1982). A New Look at the Saturn System: The Voyager 2 Images. *Science* 215, 504-537. Doi: 10.1126/science.215.4532.504.

Sromovsky, L. A., H. E. Revercomb, R. J. Krauss et al. (1983). Voyager 2 observations of Saturn's northern mid-latitude cloud features: morphology, motions, and evolution, *J. Geophys. Res.* 88, 8650–8666. Doi: 10.1029/JA088iA11p08650

Sromovsky, L.A., K. H. Baines and P. M. Fry (2013). Saturn's Great Storm of 2010–2011: evidence for ammonia and water ices from analysis of VIMS spectra. *Icarus* 226, 402–418. doi: 10.1016/j.icarus.2013.05.043.

Sromovsky, L.A., K. H. Baines and P. M. Fry (2018). Models of bright storm clouds and related dark ovals in Saturn's Storm Alley as constrained by 2008 Cassini/VIMS spectra. *Icarus* 302, 360-385.

Suggs R. and R. Beebe (1983). Saturn, *I.A.U. Circular* No. 3789.

Sugiyama K., K. Nakajima, M. Odaka et al. (2011). Intermittent cumulonimbus activity breaking the three-layer cloud structure of Jupiter. *Geophys. Res. Lett.* 38, 201-206.

Sugiyama K., K. Nakajima, M. Odaka, et al. (2014). Numerical simulations of Jupiter's moist convection layer: Structure and dynamics in statistically steady states. *Icarus* 231, 407-408.

Turman, B. N. (1977). Detection of lightning superbolts. *Journal of Geophysical Research* 82, 18, 2566-2568.

Vasavada, A. R., S. M. Hörst, M. R. Kennedy, et al. (2006). Cassini imaging of Saturn: Southern hemisphere winds and vortices. *Journal of Geophysical Research* 111, E05004. Doi: 10.1029/2005JE002563.

Warwick J. W., J. B. Pearce, D. R. Evans et al. (1981). Planetary radio astronomy observations from Voyager 1 near Saturn. *Science*, 212, 239-243, doi:10.1126/science.212.4491.239.

West R. A., K. H. Baines, E. Karkoschka et al. (2009). Clouds and Aerosols in Saturn's Atmosphere. Chapter 7 in Saturn after Cassini-Huygens. M. Dougherty, L. Esposito and T. Krimigis (edt.), Springer-Verlag, pp. 161-179.

Weidenschilling, S. J., and J. S. Lewis (1973). Atmospheric and cloud structures of the Jovian planets. *Icarus* 20, 465–476.

Westphal J. A., W. A. Baum, A. P. Ingersoll, et al. (1992). Hubble space telescope observations of the 1990 equatorial disturbance on Saturn: Images, albedos, and limb darkening. *Icarus* 100, 485-498. Doi: 10.1016/0019-1035(92)90112-K.

Zakharenko V., C. Mylostna, A. Konovalenko et al. (2012). Ground-based and Spacecraft

Observations of Lightning Activity on Saturn. *Planetary and Space Science*, 61, 53-59, doi:10.1016/j.pss.2011.07.021.

Zarka P., B. Cecconi, L. Denis, et al. (2006). Physical Properties and Detection of Saturn's Lightning Radio Bursts, in Planetary Radio Emissions VI, edited by Rucker, H. O., Kurth, W. S., and Mann, G., Austrian Academy of Sciences Press, Vienna, 111-122.

**Acknowledgments**

The UPV/EHU team (Spain) is supported by Grant PID2019-109467GB-I00 funded by 1042 MCIN/AEI/10.13039/501100011033/ and by Grupos Gobierno Vasco IT1742-22. EGM is Serra Hunter Fellow at UPC. EGM and ASL thankfully acknowledge the computer resources at Mare Nostrum and the technical support provided by Barcelona Supercomputing Center (AECT-2019-2-0006).